\documentclass[%
 aip,
 amsmath,amssymb,
 reprint,%
]{revtex4-1}

\usepackage{graphicx}
\usepackage{dcolumn}
\usepackage{bm}

\usepackage[utf8]{inputenc}
\usepackage[T1]{fontenc}
\usepackage{mathptmx}
\usepackage{etoolbox}

\usepackage[hidelinks]{hyperref}
\usepackage{amssymb}
\usepackage{amsmath}
\usepackage{multirow}
\usepackage{cleveref}

\usepackage{xcolor}
\usepackage{xargs}
\usepackage[colorinlistoftodos,textsize=footnotesize]{todonotes}

\DeclareSymbolFont{newfont}{OML}{cmm}{m}{it}
\DeclareMathSymbol{\Epsilon}{3}{newfont}{15}
\DeclareMathSymbol{\Varrho}{3}{newfont}{37}

\newcommandx{\unsure}[2][1=]{\todo[linecolor=red,backgroundcolor=red!25,bordercolor=red,#1]{#2}}
\newcommandx{\change}[2][1=]{\todo[linecolor=blue,backgroundcolor=blue!25,bordercolor=blue,#1]{#2}}
\newcommandx{\info}[2][1=]{\todo[linecolor=OliveGreen,backgroundcolor=OliveGreen!25,bordercolor=OliveGreen,#1]{#2}}

\newcommand{\ten}[1]{\ensuremath{\mathbf{#1}}}

\makeatletter
\def\@email#1#2{%
 \endgroup
 \patchcmd{\titleblock@produce}
  {\frontmatter@RRAPformat}
  {\frontmatter@RRAPformat{\produce@RRAP{*#1\href{mailto:#2}{#2}}}\frontmatter@RRAPformat}
  {}{}
}%
\makeatother
\begin{document}

\preprint{AIP/123-QED}

\title[How to train your solver: A method of manufactured solutions for
weakly-compressible smoothed particle hydrodynamics]{How to train your solver: A method of manufactured solutions for
weakly-compressible smoothed particle hydrodynamics}
\author{Pawan Negi}
 \email{pawan.n@aero.iitb.ac.in}
\author{Prabhu Ramachandran}%
 \email{prabhu@aero.iitb.ac.in}
\affiliation{Department of Aerospace Engineering, Indian Institute of
Technology Bombay, Powai, Mumbai 400076
}%

\date{\today}

\begin{abstract}
  \begin{sloppypar} \noindent The Weakly-Compressible Smoothed Particle
  Hydrodynamics (WCSPH) method is a Lagrangian method that is typically used for
  the simulation of incompressible fluids. While developing an SPH-based scheme
  or solver, researchers often verify their code with exact solutions, solutions
  from other numerical techniques, or experimental data. This typically requires
  a significant amount of computational effort and does not test the full
  capabilities of the solver. Furthermore, often this does not yield insights on
  the convergence of the solver. In this paper we introduce the method of
  manufactured solutions (MMS) to comprehensively test a WCSPH-based solver in a
  robust and efficient manner. The MMS is well established in the context of
  mesh-based numerical solvers. We show how the method can be applied in the
  context of Lagrangian WCSPH solvers to test the convergence and accuracy of
  the solver in two and three dimensions, systematically identify any problems
  with the solver, and test the boundary conditions in an efficient way. We
  demonstrate this for both a traditional WCSPH scheme as well as for some
  recently proposed second order convergent WCSPH schemes. Our code is open
  source and the results of the manuscript are reproducible.  \end{sloppypar}
\end{abstract}

\maketitle

\section{Introduction}
\label{sec:intro}

It has been more than four decades since the Smoothed Particle Hydrodynamics
(SPH) was first introduced~\cite{monaghan-gingold-stars-mnras-77,lucy77}. SPH
is a meshless method and is typically implemented using Lagrangian particles.
The method has been applied to a wide variety of
problems~\cite{liMultiphaseSmoothedParticle2021,yeSmoothedParticleHydrodynamics2019a,adepu2021}.
However, convergence of the SPH schemes is still considered a grand challenge
problem today~\cite{vacondio_grand_2020}. This is in part because of the
Lagrangian nature of the scheme. In this paper we introduce a powerful,
systematic methodology called the method of manufactured
solutions~\cite{roache1998verification} to study the accuracy and convergence
of the SPH method.

The method of manufactured solutions~\cite{roache1998verification} is a
well established method employed in the finite
volume~\cite{waltzManufacturedSolutionsThreedimensional2014,choudharyVerificationCompressibleIncompressible2015,choudharyCodeVerificationMultiphase2016}
and finite element~\cite{gfrererCodeVerificationExamples2018} method
communities to verify the accuracy of solvers. An important part of this
involves the verification of order of convergence guarantees provided by
the solver. \citet{roache1998verification} and thereafter
\citet{salariCodeVerificationMethod2000} formally introduced the idea of
verification and validation in the context of computational solvers for
PDEs. Verification is a mathematical exercise wherein we assess if the
implementation of a numerical method is consistent with the chosen
governing equations. For example, verification will allow us to check
whether the numerical implementation of a second-order accurate method is
indeed second-order. On the other hand, validation tests whether the chosen
governing equations suitably model the given physics. This is often
established by comparison with the results of experiments.

According to \citet{royReviewCodeSolution2005}, verification can be
classified into two categories namely, code verification, and solution
verification. In code verification, the code is tested for its correctness,
whereas in solution verification, we quantify the errors in the solution
obtained from a simulation. For example, in solution verification we
solve a specific problem and estimate the error through some means like a
grid convergence study. \citet{salariCodeVerificationMethod2000}
proposed different methods for code verification viz.\ trend test, symmetry
test, comparison test, \emph{method of exact solution} (MES), and the
\emph{method of manufactured solutions} (MMS).

In the context of SPH, the comparison test and the method of exact solution
are used widely to verify new schemes. In the comparison test, a solution
obtained from an experiment or a well-established solver is compared with
the solution obtained from the solver being tested. Many
authors~\cite{edac-sph:cf:2019,Adami2013,oger_ale_sph_2016,huang_kernel_2019}
use the computational results for the lid-driven cavity and flow past a
cylinder problems to demonstrate the accuracy of their respective solvers.
On the other hand, some
authors~\cite{damaziakComparisonSPHFEM2019,douillet-grellierComparisonMultiphaseSPH2019,dimascioSmoothedParticleHydrodynamics2017}
use solutions from established solvers to study the accuracy. In the MES,
the exact solution of the governing equations is used to compare the
accuracy as well as the order of convergence of the solver. For example,
some authors~\cite{edac-sph:cf:2019,Adami2013,sun2017deltaplus} use the
Taylor-Green vortex problem whereas
others~\cite{chaniotis2002remeshed,nasar2019} use the Gresho-Chan vortex
problem. We note that none of these studies have demonstrated formal
second-order convergence for the Lagrangian Weakly-Compressible SPH (WCSPH)
scheme. Recently, \citet{negi2021numerical} propose a family of
second-order convergent WCSPH schemes and employ the Taylor-Green problem
to demonstrate the convergence.

Despite their extensive use, the comparison and MES tests have several
shortcomings~\cite{salariCodeVerificationMethod2000}. The comparison test
often requires a significant amount of computation since a full simulation
for some complex problem is usually undertaken requiring a reasonable
resolution and a large number of timesteps to attain an appropriate
solution. In the case of the MES, there are very few exact solutions that
exercise the full capabilities of the solver. For example the Taylor-Green
and Gresho-Chan vortex problems are usually simulated without any solid
boundaries and are only available in two-dimensions. The problems are also
fairly simple and are for incompressible fluids and this imposes additional
constraints on WCSPH schemes which are not truly incompressible. For
example, \citet{negi2021numerical} show that the error of the WCSPH scheme
is $O(M^2)$, where $M$ is the Mach number of the flow, due to
the artificial compressibility assumption. Thus, the verification process
requires that the WCSPH solver be executed with significantly larger sound
speeds than normally employed further increasing the execution time.
Moreover, these methods cannot ensure that all the aspects of the solver
are tested for example, it is difficult to find the order of convergence of
the boundary condition implementation.

The method of manufactured solutions does not suffer from these shortcomings
and is considered a state-of-the-art method for the verification of
computational codes. However, this method has to our knowledge not been used
in the context of the SPH thus far. In the MMS, a solution $u=\phi(x, y, z,
t)$ is manufactured such that it is sufficiently complex and satisfies some
desirable properties~\cite{salariCodeVerificationMethod2000}. We discuss these
properties in detail in a later section (see \cref{sec:mms}). Let the
governing equation be given by
\begin{equation}
  \mathcal{F} u = g,
  \label{eq:dummy}
\end{equation}
where $\mathcal{F}$ is the differential operator, $u$ is the variable and $g$
is the source term. We subject the \emph{Manufactured Solution} (MS)
$u=\phi(x, y, z, t)$ to the governing differential equation in
\cref{eq:dummy}. Since $\phi$ may not be the solution of the governing
equation, we obtain a residual,
\begin{equation}
  r =  \mathcal{F} \phi - g.
  \label{eq:dummy_r}
\end{equation}
We add the residual $r$ as a source term to the governing equation therefore,
the modified equation is given by
\begin{equation}
  \mathcal{F} u = g + r.
  \label{eq:dummy_s}
\end{equation}
We then solve the problem along with this additional source term added to the
solver. If the solver is correct we should obtain the MS, $u$, as the
solution. We add the source term to each particle directly and this does not
change the solver in any other way. The convergence of the solver may be
computed numerically by solving the problem at different resolutions and
finding the error in the solution.

The MMS is therefore an elegant yet simple technique to test the accuracy
of a solver without making changes to the solver or the scheme. The only
requirement is that it be possible to add an arbitrary source term to a
particular equation. It is easy to see that the method can be applied in
arbitrary dimensions. Further, we may use this technique to also test
boundary conditions. By employing a carefully chosen MS one may use the
method to identify specific problems with certain discretizations. For
example, one may choose an inviscid solution to test only the pressure
gradient term in the momentum equation. This makes it easy to discover
issues in the implementation.

In \citet{feng2016} the authors use an MMS to verify their SPH
implementation. However, the particles do not move and therefore it is no
different than a traditional application of MMS in mesh-based methods. As
mentioned earlier, the MMS has not to our knowledge been applied in the
context of the Lagrangian SPH method in order to study its accuracy. It is
not entirely clear why this is the case but we conjecture that this is
because the SPH method is Lagrangian and the traditional MMS has been
applied in the case of traditional finite volume and finite element
methods. When the particles move, it becomes difficult to satisfy the
boundary conditions and have the particles moving in an arbitrary fashion.
However, these issues can be handled in the context of an SPH scheme since
it is possible to add and remove particles into a simulation. The lack of
second order convergent SPH schemes is also a possible reason for the lack
of adoption of the MMS in the SPH community. In the present work we use the
recently proposed second-order convergent Lagrangian SPH schemes
\cite{negi2021numerical} to demonstrate the method. We observe that in the
present work, all the schemes we consider employ some form of particle
shifting~\cite{acc_stab_xu:jcp:2009,lind2012incompressible,Adami2013,huang_kernel_2019}.
This is crucial since the particles can then be constrained inside a solid
domain and even if the particles move, their motion is corrected by the
particle shifting algorithm. We thus do not need to add or remove particles
from any of our simulations.

Our major contribution in this work is to show how one can apply the MMS to
carefully study the accuracy of a modern WCSPH implementation. We first
obtain a suitable initial particle configuration to be used in the
simulation. We then systematically show the method to construct a MS for
established WCSPH schemes as well as the second-order schemes proposed by
\citet{negi2021numerical}. We show how this can be applied to any specified
shape of the domain. We show how to apply the MMS in the context of both
Eulerian and Lagrangian SPH schemes. We then demonstrate how the MMS can be
useful to debug a solver by deliberately changing one of the equations in
the second-order convergent scheme and show the MS construction such that
the change is highlighted in the order of convergence plot. We then study
the convergence of some commonly used implementations for the Dirichlet and
Neumann boundary conditions for solids. We demonstrate that the method can
be used to study convergence for extreme resolutions as well as for three
dimensional cases. The proposed method is very fast as we do not require a
large number of iterations to verify the convergence. It is important to
note that while we focus on verification, a validation study must be
performed to ensure that the physics is accurately captured by the solver.

In summary, we present a simple, efficient, and powerful method to study
convergence, and perform code verification of a WCSPH solver. This is very
important given that the convergence of SPH schemes is still considered a
grand-challenge problem~\cite{vacondio_grand_2020}. We make our code
available as open source (\url{https://gitlab.com/pypr/mms_sph}) and all
the results shown in our work are fully automated in the interest of
reproducibility. In the next section we briefly discuss the SPH method
followed by the verification techniques used in SPH. Thereafter we discuss
the MMS method and how it can be applied in the context of the WCSPH
scheme. We then apply the method to a variety of problems.

\section{The SPH method}
\label{sec:sph}

In the present work, we discretize the domain $\Omega$ into equally spaced
points having mass $m$ and volume $\omega$. We may approximate a function $f$
at a point $\ten{x}_i$ in the domain $\Omega$ by,
\begin{equation}
  \left<f(\ten{x}_i)\right> = \sum_j f(\ten{x}_j) W_{ij} \omega_j,
  \label{eq:sph_f}
\end{equation}
where $W_{ij} = W(\ten{x}_i-\ten{x}_j, h)$, where $W$ is the smoothing kernel
and $h$ is its support radius, $\omega_j = m_j/\rho_j$, $\rho_j = \sum_j m_j
W_{ij}$ and $m_j$ is the mass of the particle. The sum $j$ is over all the
neighbor particles of the particle $i$. $\rho_j$ is commonly called the
\emph{summation density} in the SPH literature. The \cref{eq:sph_f} is
$O(h^2)$ accurate in a uniform domain with kernel having full
support~\cite{quinlan_truncation_2006,fatehi_error_2011}. In order to obtain
the gradient of the function $f$ at $\ten{x}_i$ using the kernel having full
support, one may use
\begin{equation}
  \left<\nabla f(\ten{x}_i)\right> = \sum_j (f(\ten{x}_j) - f(\ten{x}_i)) \tilde{\nabla} W_{ij} \omega_j,
  \label{eq:sph_df}
\end{equation}
where $\tilde{\nabla} W_{ij} = B_i \nabla W_{ij}$, where $B_i$ is the
Bonet-Lok correction matrix~\cite{bonet_lok:cmame:1999} and where $ \nabla
W_{ij}$ is the gradient of $W_{ij}$ w.r.t.\ $\ten{x}_i$. In a similar manner,
many
authors~\cite{monaghan1983shock,Adami2013,dilts1999moving,bonet_lok:cmame:1999,fatehi_error_2011}
propose various discretizations of the gradient, divergence, and Laplacian of
a function; these various forms are summarized and compared in
\onlinecite{negi2021numerical}.

The SPH method can be used to solve the Weakly-Compressible SPH equation given by
\begin{equation}
  \begin{split}
  \frac{d \Varrho}{dt} &= - \Varrho \nabla \cdot \ten{u}, \\
  \frac{d \ten{u}}{dt} &= - \frac{\nabla p}{\Varrho} + \nu \nabla^2 \ten{u},\\
  \end{split}
  \label{eq:ns_wc}
\end{equation}
where $\Varrho$, $\ten{u}$, and $p$ are the density, velocity, and pressure of
the flow, respectively, and $\nu$ is the dynamic viscosity of the fluid. We
note here that $\Varrho$ is different from the summation density $\rho$. We
use $\rho_j$ to estimate the particle volume, $\omega_j$. The governing
equations in \cref{eq:ns_wc} are completed by linking the pressure $p$ to
density $\Varrho$ using an equation of state. There are many different
schemes~\cite{sph:fsf:monaghan-jcp94,Adami2013,edac-sph:cf:2019,sun2017deltaplus,oger_ale_sph_2016}
that solve \cref{eq:ns_wc}. However, they all fail to show second-order
convergence. Recently, \citet{negi2021numerical} performed a convergence study
of various discretization operators, and propose a family of second-order
convergent schemes. In this paper, we use these schemes to demonstrate the new
method to study convergence of SPH schemes and compare it with the
\emph{Entropically damped artificial compressibility} (EDAC)
scheme~\cite{edac-sph:cf:2019}. We summarize the schemes considered in this
study as follows:
\begin{enumerate}
\item L-IPST-C (Lagrangian-Iterative PST-Coupled scheme), which is a second
  order scheme proposed in \onlinecite{negi2021numerical}, where we discretize the
  continuity equation as,
  \begin{equation}
    \frac{d \Varrho_i}{dt} = -\Varrho_i \sum_j (\ten{u}_j - \ten{u}_i) \cdot
    \tilde{\nabla} W_{ij} \omega_j.
    \label{eq:asym_cont}
  \end{equation}
  We discretize the momentum equation as,
  \begin{equation}
    \begin{split}
    \frac{d\ten{u}_i}{dt}= &- \sum_j \frac{(p_j - p_i)}{\Varrho_i} \tilde{\nabla}
    W_{ij} \omega_j + \\
    ~& \nu \sum_j (\left< \nabla \ten{u}\right>_j - \left< \nabla
    \ten{u}\right>_i ) \cdot \tilde{\nabla} W_{ij} \omega_j\\
    \end{split}
    \label{eq:asym_mom}
  \end{equation}
  where $\tilde{\nabla W_{ij}} = B_i \nabla W_{ij}$, where $B_i$ is the
  correction matrix~\cite{bonet_lok:cmame:1999}, and the $\left< \nabla
    \ten{u}\right>_i$ is the first order consistent gradient approximation
  given by
  \begin{equation}
    \left< \nabla \ten{u}\right>_i = \sum_j (\ten{u}_j - \ten{u}_i) \otimes \tilde{\nabla}
    W_{ij} \omega_j.
    \label{eq:asym_grad}
  \end{equation}
  In order to complete the system, we use a linear equation of state (EOS) where
  we link pressure with the fluid density $\Varrho$ given by
  \begin{equation}
    p_i = c_o^2 (\Varrho_i - \Varrho_o),
    \label{eq:linear_eos}
  \end{equation}
  where $c_o$ is the artificial speed of sound and $\Varrho_o$ is the reference
  density. We use the standard Runge-Kutta second order integrator for time
  stepping. The time step $\Delta t$ is set using the stability condition given by
  \begin{equation}
    \begin{split}
      \Delta t_{cfl} =& 0.25 \frac{h}{c_o + U},\\
      \Delta t_{viscous} = & 0.25 \frac{h^2}{\nu},\\
      \Delta t_{force} = & 0.25 \sqrt{\frac{h}{|\ten{g}|}},\\
      \Delta t =& \min(\Delta t_{cfl}, \Delta t_{viscous}, \Delta t_{force}),
    \end{split}
    \label{eq:dt}
  \end{equation}
  where $U$ is the maximum velocity in the domain, $\ten{g}$ is the magnitude of
  the acceleration due to gravity. For all over testcase, we set $c_o=20m/s$
  irrespective of the maximum velocity in the domain. After every ten time step,
  particle shifting is applied using iterative particle shifting technique (IPST)
  to redistribute the particle in order to obtain a uniform distribution. We
  perform first order Taylor-series correction for velocity, and density after
  shifting.

\item PE-IPST-C (Pressure Evolution-Iterative PST-Coupled scheme): This method
  is a variation of the L-IPST-C scheme where a pressure evolution equation is
  used instead of a continuity equation~\cite{negi2021numerical}. The pressure
  evolution equation is given by
  \begin{equation}
    \frac{dp}{dt} = - \Varrho c_o^2 \nabla \cdot \ten{u} + \nu_{edac} \nabla^2 p,
    \label{eq:pres_evol}
  \end{equation}
where $\nu_{edac} = \alpha h c_o /8$ with $\alpha=0.5$. The SPH discretization
of \cref{eq:pres_evol} is given by
  \begin{equation}
    \begin{split}
    \frac{d p}{dt} = &- \Varrho_i c_o^2 \sum_j (\ten{u}_j - \ten{u}_i) \cdot
    \tilde{\nabla} W_{ij} \omega_j +\\
    ~& \nu_{edac} \sum_j (\left< \nabla p\right>_j - \left< \nabla
    p\right>_i ) \cdot \tilde{\nabla} W_{ij} \omega_j,
    \end{split}
    \label{eq:pres_evol_sph}
  \end{equation}
where $\left< \nabla p\right>_i$ is evaluated using second-order consistent
approximation. Since the pressure is linked with density, we evaluate the
density by inverting the linear EOS given by
  \begin{equation}
    \Varrho_i = \frac{p_i}{c_o^2} + \Varrho_o.
    \label{eq:eos_invert}
  \end{equation}

\item TV-C (Transport Velocity-Coupled): In this method, we start with the
  Arbitrary Eulerian Lagrangian SPH
  equation~\cite{oger_ale_sph_2016,sun2019consistent} given by
  \begin{equation}
    \begin{split}
      \frac{\tilde{d} \Varrho}{dt} &= - \Varrho \nabla \cdot (\ten{u + \delta u}) +
      \nabla \cdot (\Varrho \delta \ten{u}), \\
      \frac{\tilde{d} \ten{u}}{dt} &= - \frac{\nabla p}{\Varrho} + \nu \nabla^2
      \ten{u} + \nabla \cdot (\ten{u} \otimes \delta \ten{u}) - \ten{u}
      \nabla(\delta \ten{u}),\\
      \end{split}
      \label{eq:wc_ale}
  \end{equation}
  where $\frac{\tilde{d}(\cdot)}{dt} = \frac{\partial (\cdot) }{\partial t} +
  (\ten{u} + \delta \ten{u}) \cdot \nabla (\cdot)$ and $\delta \ten{u}$ is the
  shifting velocity computed using
  \begin{equation}
    \delta \ten{u} = - M (2 h) c_o \sum_j \left[ 1 + R
    \left(\frac{W_{ij}}{W(\Delta s)} \right)^n\right] \nabla W_{ij} \omega_j,
    \label{eq:shift}
  \end{equation}
where $R=0.24$, and $n=4$~\cite{sph:tensile-instab:monaghan:jcp2000}.  We note
that the density $\Varrho$ is treated as a fluid property independent of
particle positions~\cite{negi2021numerical}. The main idea is to redistribute
the particles using a shifting force in the governing equations instead of
performing shifting post step. All the terms in the \cref{eq:wc_ale} are
discretized using a second-order accurate formulation as done in case of the
L-IPST-C scheme (for details refer to \onlinecite{negi2021numerical}).
\item E-C : This is an Eulerian method proposed by \citet{negi2021numerical}.
  The governing equations for the scheme is given by
\begin{equation}
  \begin{split}
    \frac{\partial \Varrho}{\partial t} =& - \Varrho \nabla \cdot \ten{u} -
    \ten{u} \cdot \nabla \Varrho,\\
    \frac{\partial \ten{u}}{\partial t} = & -\frac{\nabla p}{\Varrho} + \nu \nabla^2 \ten{u} -
    \ten{u} \cdot \nabla \ten{u}.\\
  \end{split}
  \label{eq:ec_eq2}
\end{equation}
A similar method was proposed by \citet{nasar2019}. However, unlike the E-C
method they evaluate the density as a function of particle distribution. This
assumption allowed them to set the last term in the continuity equation equal
to zero. This results in an increased error in the pressure as shown in
\onlinecite{negi2021numerical}. All the terms in the governing equations in the
\cref{eq:ec_eq2} are discretized using a second order accurate formulation as
done in case of L-IPST-C scheme.
\item EDAC: In this method, proposed by \citet{edac-sph:cf:2019}, we employ
  the pressure evolution equation; however, density is evaluated using
  summation density formulation ($\Varrho = \rho$ in \cref{eq:pres_evol}).
  Unlike the other methods considered above, this is not a second order
  accurate method. The discretization of the pressure evolution in
  \cref{eq:pres_evol} is given by
\begin{equation}
  \frac{dp_i}{dt} = \sum_j \frac{m_j \rho_i}{\rho_j} c_o^2 (\ten{u}_i -
  \ten{u}_j) \cdot \nabla W_{ij}.
  \label{eq:edac}
\end{equation}
The momentum equation is discretized as
\begin{equation}
  \begin{split}
  \frac{d u_i}{dt} = \frac{1}{m_i} \sum_j (V_i^2 + V_j^2) \left[ \tilde{p}_{ij}
  \nabla W_{ij} + \tilde{\eta}_{ij} \frac{(\ten{u}_i - \ten{u}_j)}{r_{ij}^2 +
  \eta h_{ij}^2} \nabla W_{ij} \cdot \ten{r}_{ij} \right],
  \end{split}
\end{equation}
where $\tilde{p}_{ij} = \frac{\rho_j p_i + \rho_i p_j}{\rho_i + \rho_j}$, and
$\tilde{\eta}_{ij} = \frac{2 \eta_i \eta_j}{\eta_i + \eta_j}$, where $\eta_i =
\rho_i \nu_i$.
\end{enumerate}

In the next section, we consider the standard approach employed in most SPH
literature where a code verification is performed to verify the SPH method.

\section{Code verification in SPH}
\label{sec:sph_conv}

Verification and validation of a numerical method are equally important.
Verification of the accuracy and convergence of a solver is found using
exact solutions, solutions from existing solvers, experimental results, or
manufactured solutions. The verification can also be used to identify bugs
in the solver. On the other hand, validation ensures that the governing
equations are appropriate for the physics and often involves comparison
with experimental results.

Verification is of two kinds: (i) \emph{code verification}, where we test
the code of the numerical solver for correctness and accuracy, and (ii)
\emph{solution verification}, where we quantify the error in a solution
obtained. In this paper, we focus on the code verification techniques
applied to SPH. The different techniques for code
verification~\cite{salariCodeVerificationMethod2000} are:
\begin{itemize}
\item Trend test: Where we use an \emph{expert judgment} to verify the
solution obtained. For example, the velocity of the vortex in a viscous periodic
domain should diminish with time. If the solver shows an increase of the
velocity in the domain, then there is an error in the solver.
\item Symmetry test: Where we ensure that the solution obtained does not change
if the domain is rotated or translated. For example, if we implement an inlet
assuming the flow in the $x$ direction, we will get an erroneous result on
rotating the domain by $90$ degree.
\item Comparison test: Where we compare the solution obtained from the
solver with the solutions from an established solver or experiment. This method
has been used widely by many authors in the SPH community~\cite{edac-sph:cf:2019,Adami2013,oger_ale_sph_2016,huang_kernel_2019,damaziakComparisonSPHFEM2019,douillet-grellierComparisonMultiphaseSPH2019}
to show the correctness of their respective works.
\item Method of exact solution (MES): Where we solve a problem for which the
exact solution is known. For examples, in \onlinecite{negi2021numerical} this method
is applied to the Taylor-Green problem for which an exact solution is known.
Some authors~\cite{fatehi_error_2011,schwaigerImplicitCorrectedSPH2008} use
exact solution for 1D and 2D conduction problems to demonstrate
convergence.
\end{itemize}

In the context of SPH, out of the above mentioned methods comparison test
and MES are employed widely. We compare solutions for the Taylor-Green and
lid-driven cavity problems which are the examples of MES and comparison
test, respectively.

The Taylor-Green problem has an exact solution given by
\begin{equation}
  \begin{split}
    u &= - U e^{bt} \cos(2 \pi x) \sin(2  \pi y),\\
    v &= U e^{bt} \sin(2 \pi x) \cos(2  \pi y),\\
    p &= -0.25 U^2 e^{2bt} (\cos(4 \pi x) +\cos(4  \pi y)),\\
  \end{split}
  \label{eq:tg_exact}
\end{equation}
where $b = -8 \pi^2 / Re$, where $Re$ is the Reynolds number of the flow.
We consider $Re=100$ and $U = 1m/s$. We solve this problem
for three different resolutions viz.\ $50\times50$, $100\times100$, and
$200\times200$ for a two-dimensional domain of size $1m \times 1m$ for $2$
sec using L-IPST-C scheme. However, we discretize the pressure gradient
using the formulation given by
\begin{equation}
  \left< \frac{\nabla p}{\Varrho} \right> = \sum_j \frac{(p_j + p_i)}{\Varrho_i}
  \tilde{\nabla} W_{ij} \omega_j
  \label{eq:pres_sym}
\end{equation}
\begin{figure}[htbp]
  \centering
  \includegraphics[width=\linewidth]{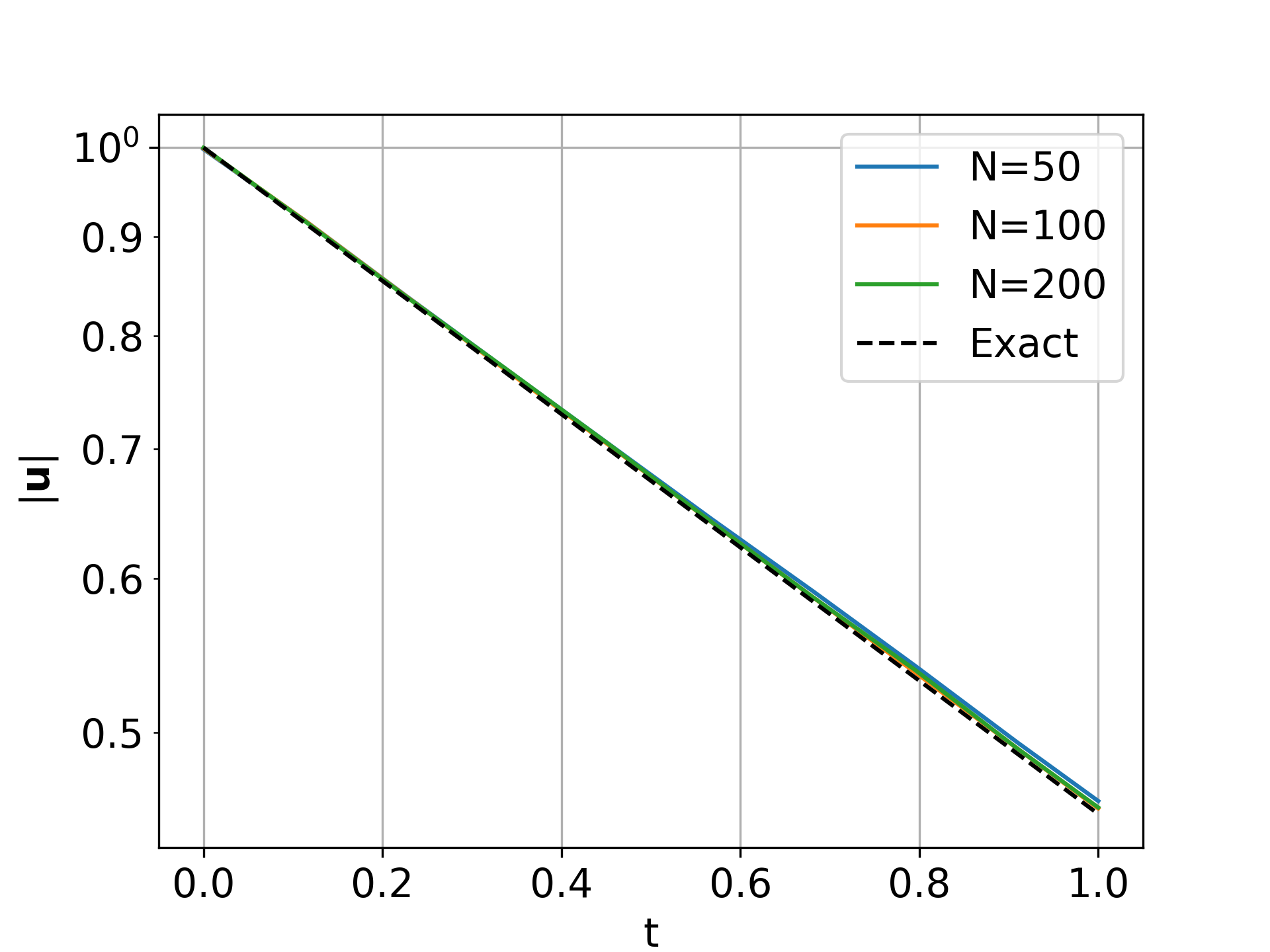}
  \caption{The decay in velocity magnitude for different resolutions
  compared with the exact solution for the Taylor-Green problem.}
  \label{fig:tg}
\end{figure}
In \cref{fig:tg}, we plot the decay in the velocity magnitude with time for
different resolution compared with the exact solution. Clearly, the
decay in the velocity magnitude is very close to the expected result.

In the lid-driven cavity problem, we consider a two-dimensional domain of
size $1m \times 1m$ with 5 layers of ghost particles representing the solid
particles. The top wall at $y=1m$ is given a velocity $u=1m/s$ along the
x-direction. We solve the problem using the L-IPST-C scheme for different
resolution for $10$ sec. However, we discretize the viscous term using the
method given by \citet{cleary1999conduction}.
\begin{figure}[htbp]
  \centering
  \includegraphics[width=\linewidth]{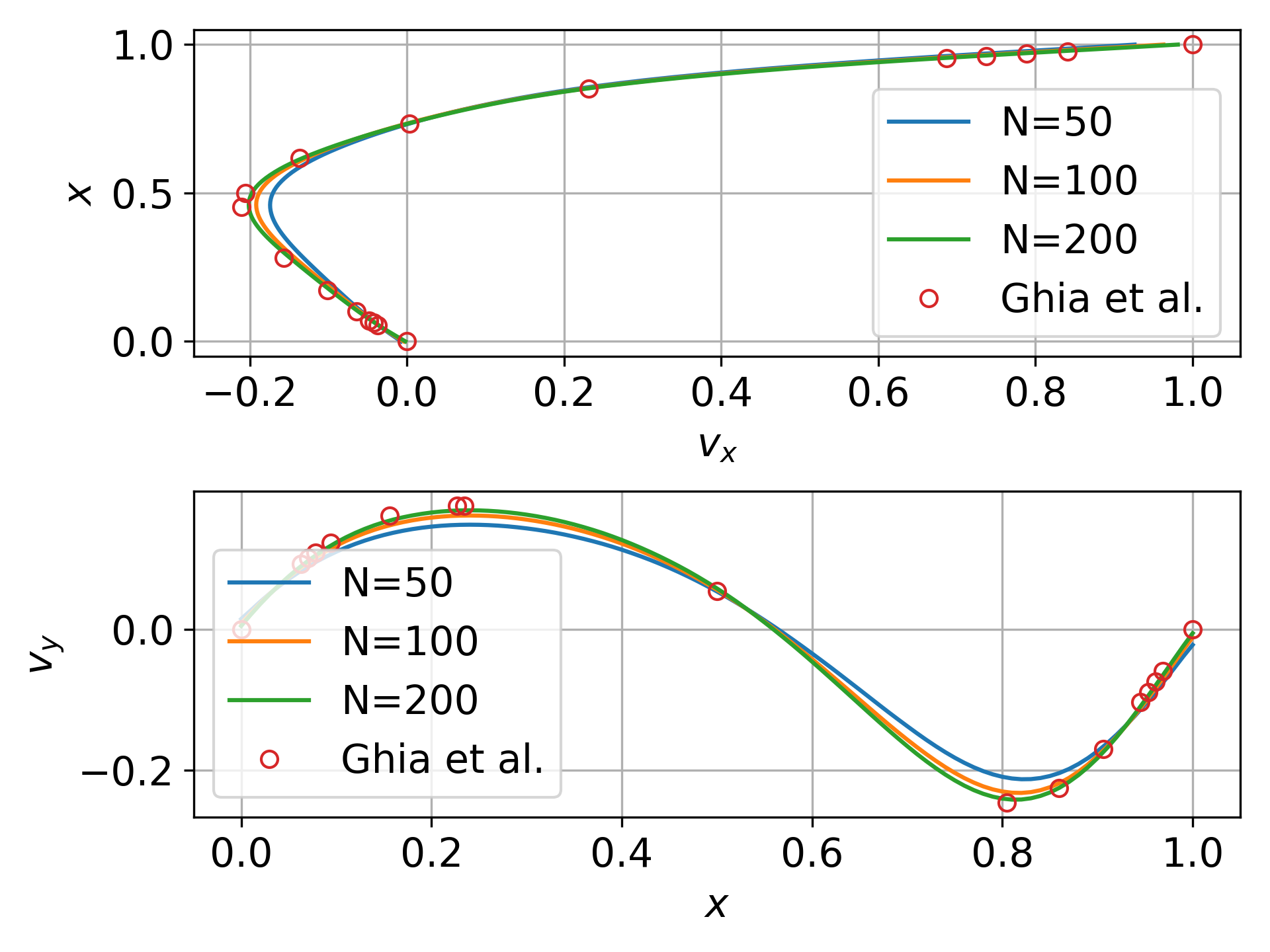}
  \caption{The velocity along $x$ and $y$ direction along the center line
  $x=0.5$ of the domain for the lid-driven cavity problem}
  \label{fig:ldc}
\end{figure}
In \cref{fig:ldc}, we plot the velocity along the centerline $x=0.5$ of the
domain compared with the result of \citet{ldc:ghia-1982}. Clearly, the
increase in resolution improves the accuracy.

We note that many
researchers~\cite{edac-sph:cf:2019,Adami2013,oger_ale_sph_2016,huang_kernel_2019,damaziakComparisonSPHFEM2019,douillet-grellierComparisonMultiphaseSPH2019}
use the above approach to verify their SPH schemes. Unfortunately, in both
problems discussed above we used a discretization which is not second-order
accurate. Evidently, these kind of verification techniques are unable to
detect such issues. In addition, the simulations take a significant amount
of time. For example, the $200\times200$ resolution lid-driven cavity case
took 150 minutes. In the case of the Taylor-Green problem since the exact
solution is known one can evaluate the $L_1$ error in velocity or pressure.
\begin{figure}[htbp]
  \centering
  \includegraphics[width=\linewidth]{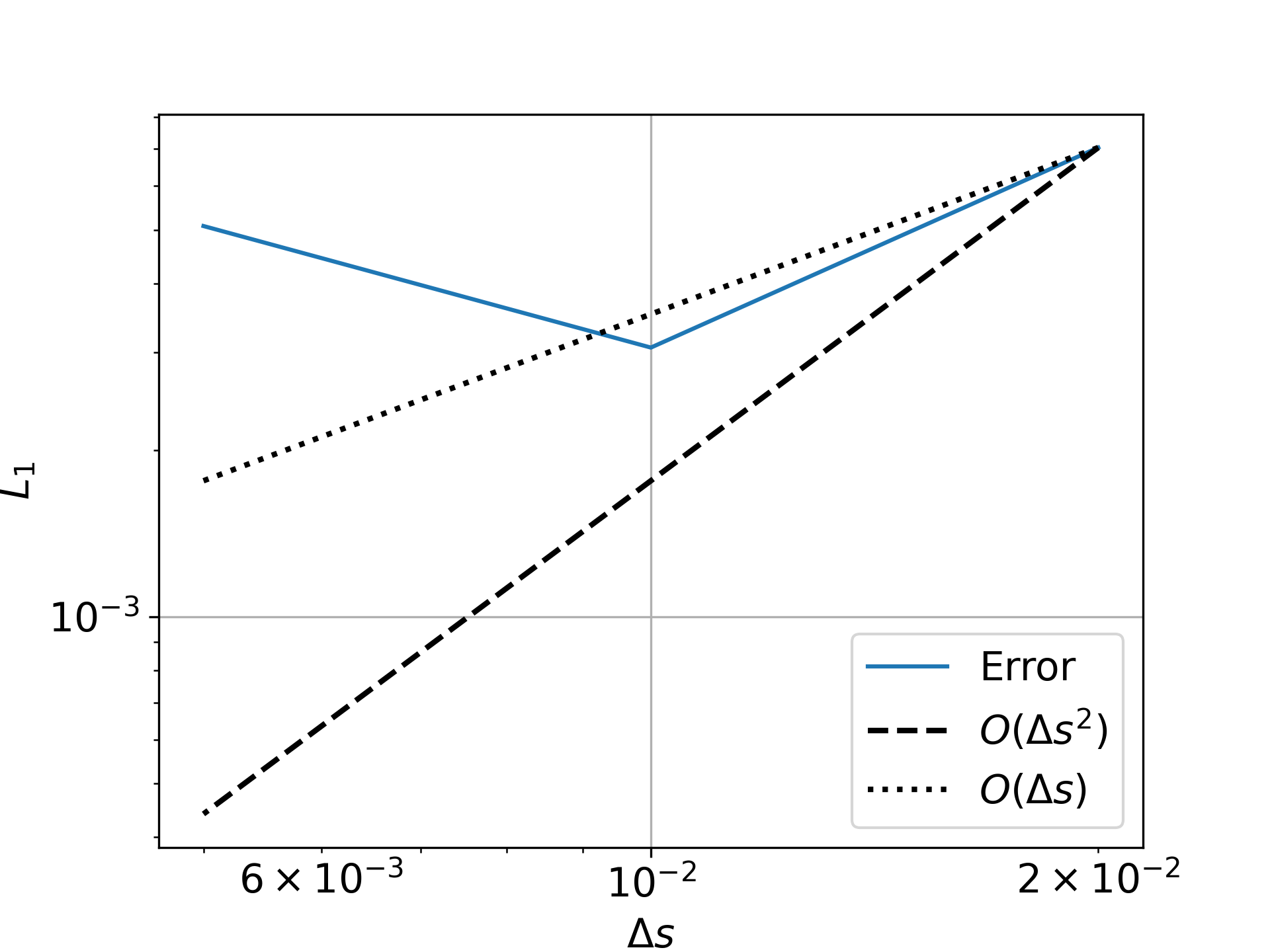}
  \caption{The $L_1$ error in velocity for the Taylor-Green problem.}
  \label{fig:l1_tg}
\end{figure}
In \cref{fig:l1_tg}, we plot the $L_1$ error in velocity as a function of
particle spacing. The $L_1$ error is not second-order and diverges as we
increase resolution from $100\times100$ to $200\times200$. However, this
result does not suggest to us the exact reason for the error.

In general, one cannot exercise specific terms in the governing
differential equation (GDE) in all the methods described above. Therefore,
the source of error cannot be determined. For example, the solver may show
convergence in the case of the Gresho-Chan vortex problem but fail for the
Taylor-Green vortex problem due to an issue with the discretization of the
viscous term. It is only
recently~\cite{antuonoTriperiodicFullyThreedimensional2020} that an
analytic solution for three dimensional Navier-Stokes equations has been
proposed. Other recent
work~\cite{sharmaVorticityDynamicsThreedimensional2019} has only focused on
numerical investigation. It is therefore difficult to apply the MES in
three dimensions. Furthermore, such studies require an even larger
computational effort. Finally, we note that the Taylor-Green vortex problem
is for an incompressible fluid making it difficult to test a WCSPH scheme.

Therefore, in the context of SPH, the comparison and MES techniques are
insufficient and inefficient. We require a better method to verify the
solver before proceeding to validation. The method of manufactured
solutions offers exactly such a technique and this is described in the next
section.

\section{The Method of Manufactured Solutions}
\label{sec:mms}

In conventional finite volume and finite element schemes, it is mandatory
to demonstrate the order of convergence and the MMS has been used for
this~\cite{choudharyCodeVerificationBoundary2016,gfrererCodeVerificationExamples2018,waltzManufacturedSolutionsThreedimensional2014}.
For the SPH method, obtaining second-order convergence has itself been a
challenge~\cite{vacondio_grand_2020} until
recently~\cite{negi2021numerical}. Moreover, to the best of our knowledge
the MMS method has not been applied in the context of SPH. In this paper,
we apply the principles of MMS to formally verify SPH solvers in a fast and
reliable manner. The technique facilitates a careful investigation of the
the various discretization operators, the boundary condition
implementation, and time integrators.

In MMS, an \emph{artificial or manufactured solution} is assumed. Let us
assume the manufactured solution (MS) for $\Varrho$, $\ten{u}$, and $p$ in
\cref{eq:ns_wc} are $\tilde{\Varrho}$, $\tilde{\ten{u}}$, and $\tilde{p}$,
respectively. Since the MS is not the solution of the \cref{eq:ns_wc}, we
obtain a residue,
\begin{equation}
  \begin{split}
  s_\Varrho &= \frac{d \tilde{\Varrho}}{dt} + \tilde{\Varrho} \nabla \cdot
  \ten{\tilde{u}}, \\
  \ten{s}_\ten{u} &= \frac{d \ten{\tilde{u}}}{dt} + \frac{\nabla
  \tilde{p}}{\tilde{\Varrho}} - \nu \nabla^2 \ten{\tilde{u}},\\
  \end{split}
  \label{eq:mms_eg}
\end{equation}
where $s_\Varrho$ and $\ten{s}_\ten{u}$ are the residue term for continuity and
momentum equation, respectively. Since, we have the closed form expression for
all the terms in the RHS of the \cref{eq:mms_eg}, we may introduce the residue
terms as source terms in the governing equations. We write the modified
governing equations as
\begin{equation}
  \begin{split}
  \frac{d \Varrho}{dt} &= - \Varrho \nabla \cdot \ten{u} + s_\Varrho, \\
  \frac{d \ten{u}}{dt} &= - \frac{\nabla p}{\Varrho} + \nu \nabla^2 \ten{u} +
  \ten{s}_\ten{u}.\\
  \end{split}
  \label{eq:ns_wc_mod}
\end{equation}
Finally, we solve the \cref{eq:ns_wc_mod}. The addition of the source terms
ensures that the solution is $\tilde{\Varrho}$, $\tilde{\ten{u}}$, and
$\tilde{p}$.

One must take few precautions while employing the
MMS~\cite{salariCodeVerificationMethod2000}:
\begin{enumerate}
  \item The MS must be $C^n$ smooth where $n$ is the order of the governing equations.
  \item It must exercise all the terms i.e., for any evolution
  equation the MS cannot be time-independent.
  \item The MS must be bounded in the domain of interest. For example, the MS
  $u=tan(x)$ in the domain $[-\pi, \pi]$ is not bounded thus, should not be used.
  \item The MS should not prevent the successful completion of the code. For
  example, if the code assumes the solution to have positive pressure, then the
  MS must make sure that the pressure is not negative.
  \item The MS should make sure that the solution satisfies the basic physics.
  For example, in a shear layer flow with discontinuous viscosity, the flux must
  be continuous.
\end{enumerate}
We note that the MS may not be physically realistic.

We modify the basic steps for MMS proposed by \citet{oberkampf2010verification}
for use in the context of WCSPH as follows:
\begin{enumerate}
\item Obtain the modified form of the governing equations as employed in the
scheme. For example, in case of the $\delta$-SPH
scheme~\cite{antuono-deltasph:cpc:2010}, the continuity equation used is,
  \begin{equation}
    \frac{d \Varrho}{dt} = -\Varrho \nabla \cdot \ten{u} + D \nabla^2 \Varrho,
    \label{eq:delta_sph}
  \end{equation}
where $D=\delta h c_o$ is the damping used, and $\delta$ is a numerical
parameter. The additional diffusive term in \cref{eq:delta_sph} must be retained
while obtaining the source term.
  \item Construct the MS using analytical functions. The general form of MS is
  given by
  \begin{equation}
    f(x, y, t) = \phi_o + \phi(x, y, t),
    \label{eq:gen_ms}
  \end{equation}
  where $f$ is any property viz.\ $\Varrho$, $\ten{u}$, or $p$; $\phi_o$ is a
  constant, and $\phi(x, y, t)$ is a function chosen such that the five
  precautions listed above are satisfied.
  \item Obtain the source term as done in \cref{eq:mms_eg}.
\item Add the source term in the solver appropriately. In SPH, the source term
$s = s(x, y, z, t)$, is discretized as $s_i=s(x_i, y_i, z_i, t)$ where subscript
$i$ denotes the $i$\textsuperscript{th} particle.
\item Solve the modified equations using the solver for different particle
spacings/smoothing length ($h$). The properties on the boundary particles are updated using the MS. We
note that in the context of WCSPH schemes, one should not evaluate the derived
quantities like gradient of velocity using the MS on the solid boundary.
  \item Evaluate the discretization error for each resolution. We evaluate
  the error using
  \begin{equation}
    L_1(h) = \sum_j \sum_i \frac{|f(\ten{x}_i, t_j) - f_o(\ten{x}_i, t_j)|}{N} \Delta t,
    \label{eq:l1_time_avg}
  \end{equation}
  where $f$ is the property of interest, $N$ is the total number of particles and
  $\Delta t$ is the time interval between consecutive solution instances.
\item Compute the order of accuracy and determine whether the desired order is
achieved.
\end{enumerate}

The solver involves discretization of the governing equations and appropriate
implementation of the boundary conditions. The MMS can be used to determine the
accuracy of both. However, to obtain the accuracy of boundary conditions, the
order of convergence of the governing equations should be at least as large as
that of the boundary conditions\cite{choudharyCodeVerificationMultiphase2016}.
\citet{bond2007manufactured} and
\citet{choudharyVerificationCompressibleIncompressible2015} proposed a method to
construct MS for boundary condition verification. In order to obtain a MS for a
boundary surface given as $F(x, y, z)=C$, we multiply the original MS with $(C-F(x,
y, z))^m$. We write the new MS as
\begin{equation}
  f_{BC}(x, y, t) = \phi_o + (C-F(x, y, z))^m  \phi(x, y, t),
\end{equation}
where $m$ is the order of the boundary condition. For example, for the Dirichlet
boundary $m=1$ and for Neumann boundary $m=2$.

In the next section, we demonstrate the application of MMS to obtain the order
of convergence for the schemes listed in \cref{sec:sph}.

\section{Results}
\label{sec:results}

In this section, we apply the MMS to obtain the order of convergence of
various schemes along with their boundary conditions. We first determine
the initial particle configuration viz.\ unperturbed, perturbed, or
packed~\cite{negi2019improved} required for the MMS. We then demonstrate
that one can apply the MMS to arbitrarily-shaped domains. We then compare
the EDAC and PE-IPST-C schemes which differ in the treatment of the
density. We next apply the MMS to E-C and TV-C schemes as they employ
different governing equations compared to standard WCSPH in
\cref{eq:ns_wc}. We also demonstrate the application of the MMS method as a
technique to identify mistakes in the implementation. Finally, we employ
the MMS to obtain the order of convergence of solid wall boundary
conditions. We consider the boundary condition proposed by
\citet{maciaTheoreticalAnalysisNoSlip2011} for the demonstration.

In all our test cases, we use the quintic spline kernel with $h_{\Delta s} =
h/{\Delta s}=1.2$, where $\Delta s$ is the initial inter-particle spacing. We
consider a domain of size $1m \times 1m$. We simulate all the test cases for $50
\times 50$, $100 \times 100$, $200 \times 200$, $250 \times 250$, $400 \times
400$, $500 \times 500$, and $1000 \times 1000$ resolutions to obtain the order
of convergence plots. In all our simulations, we initialize the particles
properties using the MS. We then solve \cref{eq:ns_wc_mod} and set the
properties on any solid particle using the MS before every timestep. We set a
fixed time step corresponding to the highest resolution for all the other
resolutions. The appropriate time step is chosen using the criteria in
\cref{eq:dt}. We evaluate the $L_1$ error using \cref{eq:l1_time_avg} in the
solution.

The implementation of the code for the source terms (as shown in
\cref{eq:mms_eg}) due to the MS are automatically generated using the
\texttt{sympy}~\cite{10.7717/peerj-cs.103} and \texttt{mako}\cite{soft:mako}
packages. We recommend this approach to avoid mistakes during implementation.
\citet{salariCodeVerificationMethod2000} used a similar approach to
automatically generate the source term for their solvers.  We use the
\texttt{PySPH}~\cite{pysph2020} framework for the implementation of the schemes
described in this manuscript. All the figures and plots in this manuscript are
reproducible with a single command through the use of the
\texttt{automan}~\cite{pr:automan:2018} framework. The source code is available
at \url{https://gitlab.com/pypr/mms_sph}.

\subsection{The effect of initial particle configuration}
\label{sec:config}

\begin{figure}[ht!]
  \centering
  \includegraphics[width=\linewidth]{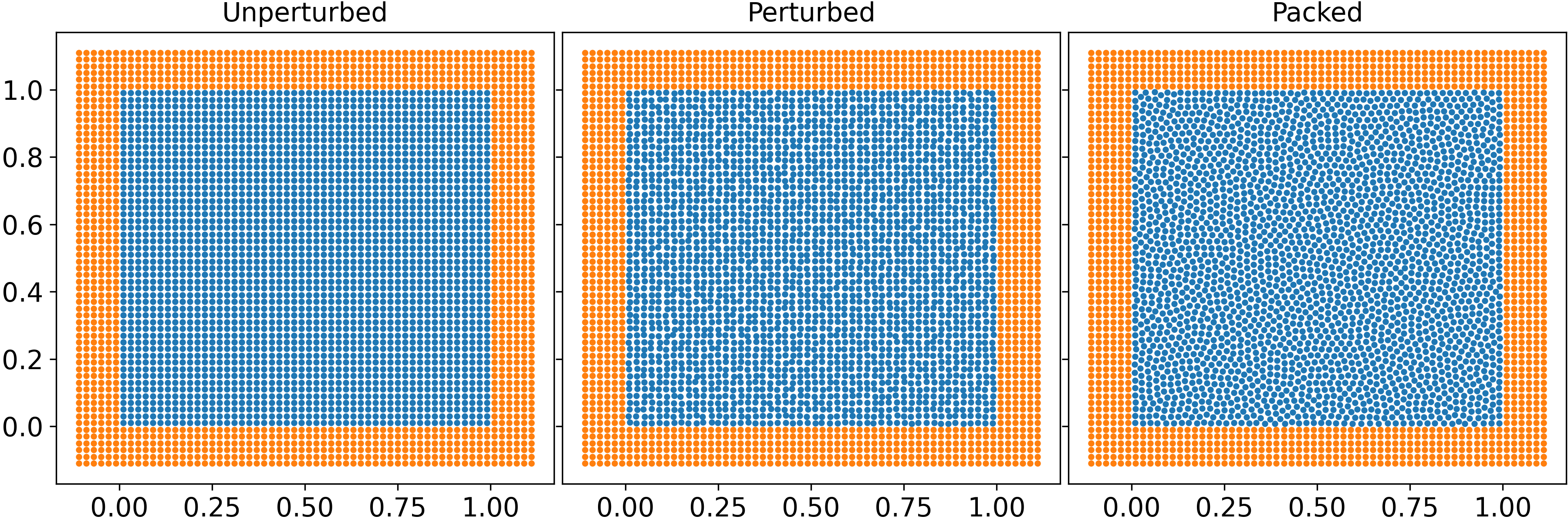}
  \caption{The different initial particle arrangements in blue with the solid
  boundary in orange.}
  \label{fig:diff_pack}
\end{figure}

The initial particle configuration plays a significant role in the error
estimation since the divergence of the velocity is captured accurately when the
particles are uniformly arranged~\cite{negi2021numerical}. In this test case, we
consider three different initial configurations of particles, widely used in SPH
literature viz.\ unperturbed, perturbed, and packed. The unperturbed
configuration is the one where we place the particles on a Cartesian grid such
that the particles are at a constant distance along the grid lines. In the
perturbed configuration, we perturb the particles initially placed on a
Cartesian grid by adding a uniformly distributed random displacement as a
fraction of the inter-particle spacing $\Delta s$. For the packed configuration,
we use the method proposed in \cite{colagrossi2012particle,negi2021numerical} to
resettle the particles from a randomly perturbed distribution to a new
configuration such that the number density of the particles is nearly constant.
In \cref{fig:diff_pack}, we show all the initial particle distributions with the
solid boundary particles in orange.

We consider the MS of the form
\begin{equation}
  \begin{split}
    u(x, y, t) =& e^{- 10 t} \sin{\left(2 \pi x \right)} \cos{\left(2 \pi y \right)}\\
    v(x, y, t) =& - e^{- 10 t} \sin{\left(2 \pi y \right)} \cos{\left(2 \pi x \right)}\\
    p(x, y, t) =& e^{-10 t} \left(\cos{\left(4 \pi x \right)} + \cos{\left(4 \pi y \right)}\right)\\
    \Varrho(x, y, t) =& \frac{p}{c_o^2} + \Varrho_o\\
  \end{split}
  \label{eq:mms_par_dist}
\end{equation}
where, we set $c_o=20 m/s$ for all our testcases. The MS complies with all the
required conditions discussed in \cref{sec:mms}. We note that the MS chosen
resembles the exact solution of the Taylor-Green problem. However, since the
solver simulates the NS equation using a weakly compressible formulation, we
obtain additional source terms when we substitute the MS to \cref{eq:ns_wc} with
$\nu=0.01 m^2/s$. We obtain the source terms from the symbolic framework,
\texttt{sympy} as,
\begin{widetext}
\begin{equation}
  \begin{split}
    s_u(x, y, t) =& 2 \pi u e^{- 10 t} \cos{\left(2 \pi x \right)} \cos{\left(2
    \pi y \right)} - 2 \pi v e^{- 10 t} \sin{\left(2 \pi x \right)} \sin{\left(2
    \pi y \right)} - 10 e^{- 10 t} \sin{\left(2 \pi x \right)} \cos{\left(2 \pi
    y \right)} + \\
    ~& 0.08 \pi^{2} e^{- 10 t} \sin{\left(2 \pi x \right)}
    \cos{\left(2 \pi y \right)} - \frac{4 \pi e^{- 10 t} \sin{\left(4 \pi x
    \right)}}{\Varrho},\\
    s_v(x, y, t) =& 2 \pi u e^{- 10 t} \sin{\left(2 \pi x \right)} \sin{\left(2
    \pi y \right)} - 2 \pi v e^{- 10 t} \cos{\left(2 \pi x \right)} \cos{\left(2
    \pi y \right)} - 0.08 \pi^{2} e^{- 10 t} \sin{\left(2 \pi y \right)}
    \cos{\left(2 \pi x \right)} + \\
    ~&10 e^{- 10 t} \sin{\left(2 \pi y \right)}
    \cos{\left(2 \pi x \right)} - \frac{4 \pi e^{- 10 t} \sin{\left(4 \pi y
    \right)}}{\Varrho},\\
    s_\Varrho(x, y, t) =& - \frac{4 \pi u e^{- 10 t} \sin{\left(4 \pi x
    \right)}}{c_{0}^{2}} - \frac{4 \pi v e^{- 10 t} \sin{\left(4 \pi y
    \right)}}{c_{0}^{2}} - \frac{10 \left(\cos{\left(4 \pi x \right)} +
    \cos{\left(4 \pi y \right)}\right) e^{- 10 t}}{c_{0}^{2}}.\\
  \end{split}
  \label{eq:source_part_dist}
\end{equation}
\end{widetext}
We add $\ten{s}_\ten{u}=s_u \hat{\ten{i}} + s_v \hat{\ten{j}}$ to the momentum
equation and $s_\Varrho$ to the continuity equation as shown in
\cref{eq:ns_wc_mod}. We solve the modified WCSPH equations in
\cref{eq:ns_wc_mod} using the L-IPST-C method for 100 timesteps where we initialize
the domain using \cref{eq:mms_par_dist}. The values of the properties $\ten{u}$,
$p$, and $\Varrho$ on the (orange) solid particles are set using
\cref{eq:mms_par_dist} at the start of every time step.

\begin{figure}[ht!]
  \centering
  \includegraphics[width=\linewidth]{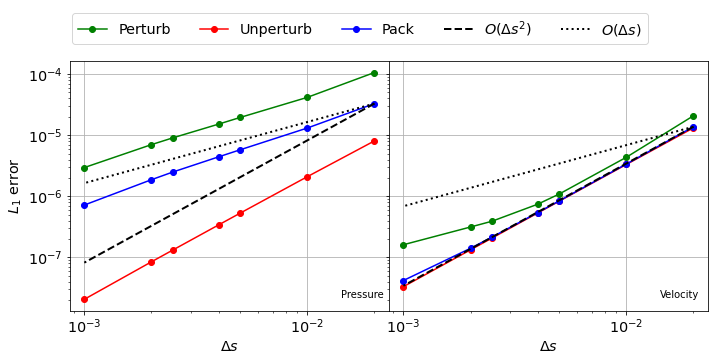}
  \caption{The error in pressure (left) and velocity (right) with fluid particles
  initialized using the MS in \cref{eq:mms_par_dist} and the source term in
  \cref{eq:source_part_dist} after 10 timesteps for the different
  configurations.}
  \label{fig:diff_pack_10}
\end{figure}

In \cref{fig:diff_pack_10}, we plot the $L_1$ error in pressure and velocity
after 10 timesteps as a function of resolution for different initial particle
distributions. Clearly, the difference in initial configuration affects the
error in pressure by a large amount. However, in velocity, the error is large in
the case of the perturbed configuration only. The unperturbed configuration has
zero divergence error at $t=0$~\cite{negi2021numerical}. Whereas, the perturbed
configuration has high error due to the random initialization. Over the course of
a few iterations, there is no significant difference between the distribution of
particles for the unperturbed and the packed configurations. Therefore, we
simulate the problems for 100 timesteps for a fair comparison.

\begin{figure}[ht!]
  \centering
  \includegraphics[width=\linewidth]{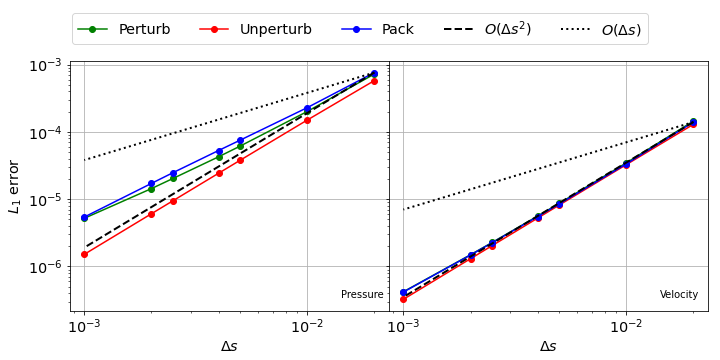}
  \caption{The error in pressure (left) and velocity (right) with fluid particles
  initialized using the MS in \cref{eq:mms_par_dist} and the source term in
  \cref{eq:source_part_dist} after 100 timesteps for all the
  configurations.}
  \label{fig:diff_pack_100}
\end{figure}

In \cref{fig:diff_pack_100}, we plot the $L_1$ error in pressure and velocity
after 100 timesteps as a function of resolution for the cases considered.
Clearly, the difference in error is reduced. However, the order of convergence
is not captured accurately. This is because the initial divergence is not
captured accurately by the packed and perturbed configurations. This difference
can be avoided through the use of a non-solenoidal velocity field. Therefore we
consider the following modified MS,

\begin{equation}
  \begin{split}
    u(x, y, t) =& y^{2} e^{- 10 t} \sin{\left(2 \pi x \right)} \cos{\left(2 \pi y
    \right)}\\
    v(x, y, t =& - e^{- 10 t} \sin{\left(2 \pi y \right)} \cos{\left(2 \pi x
    \right)}\\
    p(x, y, t) = &\left(\cos{\left(4 \pi x \right)} + \cos{\left(4 \pi y
    \right)}\right) e^{- 10 t}\\
  \end{split}
  \label{eq:mms_vis_no_div}
\end{equation}

\begin{figure}[ht!]
  \centering
  \includegraphics[width=\linewidth]{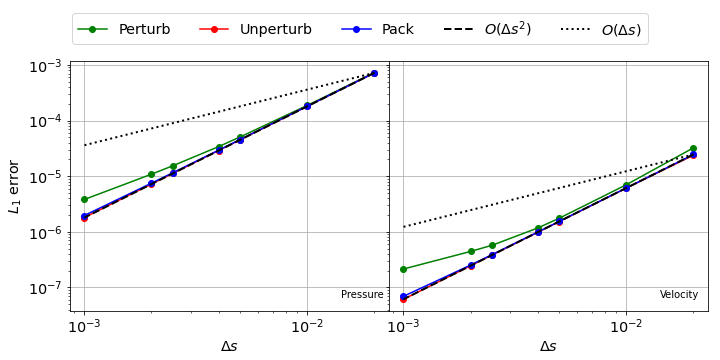}
  \caption{The error in pressure (left) and velocity (right) with fluid particles
  initialized using the MS in \cref{eq:mms_vis_no_div} and the corresponding source
  terms after 100 timesteps for all the configurations.}
  \label{fig:diff_pack_100_div}
\end{figure}

We note that the new MS velocity field is not divergence-free. We obtain the
source term with $\nu=0.01 m^2/s$ as done in \cref{eq:source_part_dist}. We
simulate the problem by initializing the domain using MS in
\cref{eq:mms_vis_no_div}. We also update the solid boundary properties using
this MS before every timestep. In \cref{fig:diff_pack_100_div}, we plot the
$L_1$ error for pressure and velocity as a function of resolution. Clearly, both
the packed and unperturbed domain show second-order convergence. Whereas, the
perturbed configuration fails to show second-order convergence. Therefore, in
the context of WCSPH schemes, one should not use a divergence-free field in the
MS. Furthermore, one should use either a packed or unperturbed configuration for
the convergence study.

It is important to note that in stark contrast the Taylor-Green vortex problem
the method shows second-order convergence irrespective of the value of $c_o$. In
\citet{negi2019improved} a much higher $c_o=80m/s$ was necessary in order to
demonstrate second-order convergence. Furthermore, the convergence is
independent of the initial configuration after 100 steps; therefore, we
recommend simulating all the testcases for at least 100 timesteps to obtain the
true order of convergence. It is important to note that some discretizations are
second-order accurate when an unperturbed configuration is
used~\cite{negi2021numerical}. In order to test the robustness of the
discretization we recommend using a packed configuration.

\subsection{The selection of the domain shape}
\label{sec:domain}

\begin{figure}[htbp]
  \centering
  \includegraphics[width=0.8\linewidth]{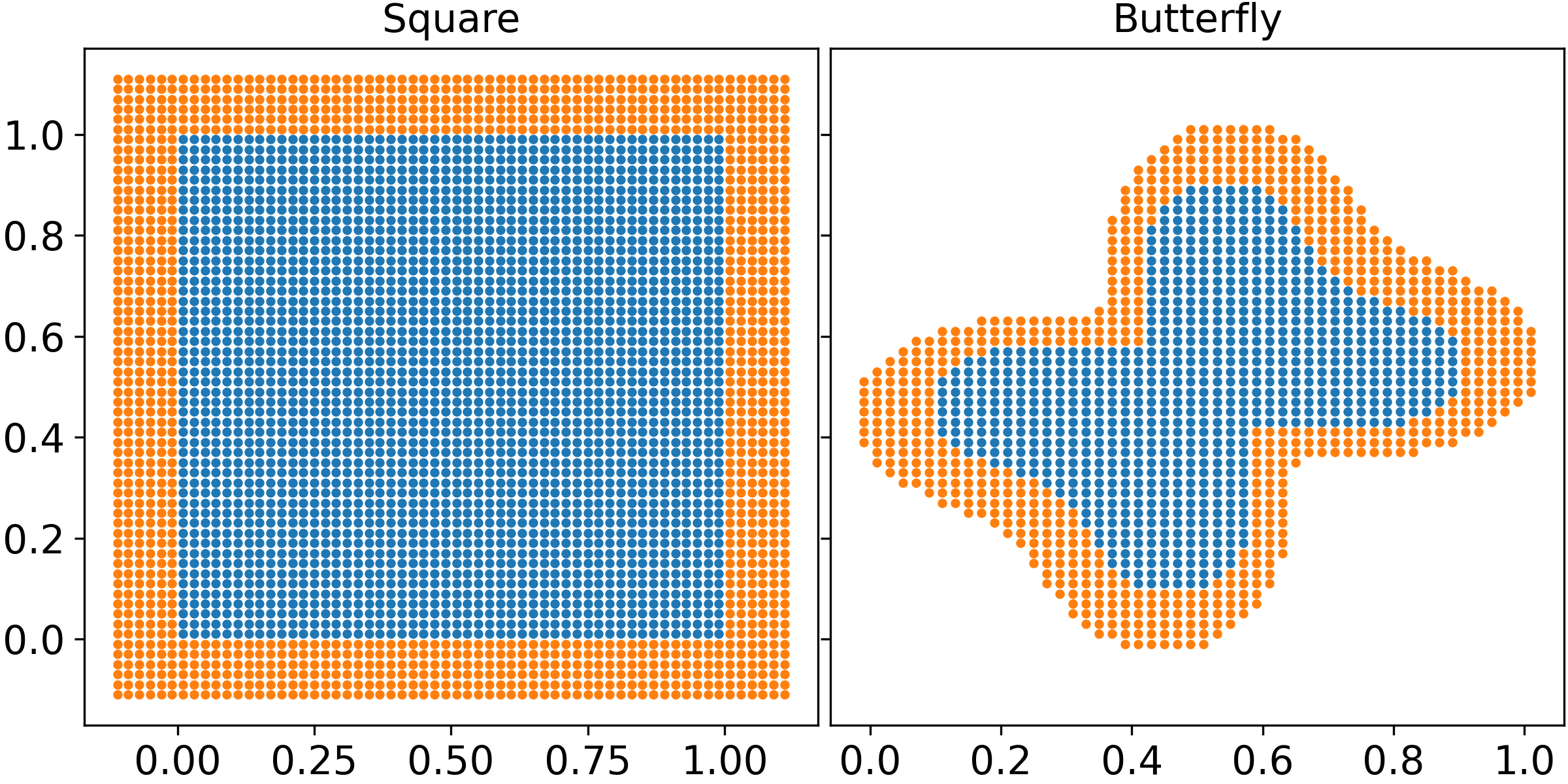}
  \caption{The different domain shapes with solid particles in orange and fluid
  particles in blue.}
  \label{fig:domain}
\end{figure}

We now show the effect of the shape of the domain on the convergence of a
scheme. We consider a square-shaped and a butterfly-shaped domain as shown in
\cref{fig:domain}.

\begin{figure}[htbp]
  \centering
  \includegraphics[width=\linewidth]{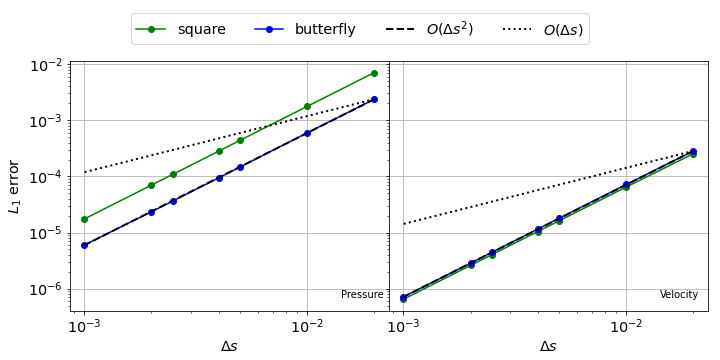}
  \caption{The $L_1$ error in pressure (left) and velocity (right) with increase
  in resolution for different shapes of the domain.}
  \label{fig:domain_conv}
\end{figure}

We consider the MS with the non-solenoidal velocity field in
\cref{eq:mms_vis_no_div} as used in the previous testcase. The source terms
obtained remains same as before, where we consider $\nu=0.01 m^2/s$. We solve
the modified equations using the L-IPST-C scheme for 100 time step for each
domain. We initialize the fluid and solid particles using the MS in
\cref{eq:mms_vis_no_div}. We update the properties of the solid particles before
every timestep using the same MS.

In \cref{fig:domain_conv}, we show the convergence of $L_1$ error after 100
timesteps in pressure and velocity as a function of resolution for both the
domain considered. Clearly, both the domains considered show second-order
convergence. Hence, one can consider any shape of the domain for the convergence
study of WCSPH schemes using MMS. However, we only use square-shaped domain for
all our test cases.

\subsection{Comparison of EDAC and PE-IPST-C}
\label{sec:edac_pe}

In this testcase, we compare the convergence of EDAC~\cite{edac-sph:cf:2019} and
PE-IPST-C~\cite{negi2021numerical} schemes. These two schemes have two major
differences. First, the discretizations used in PE-IPST-C method are all
second-order accurate in contrast to the EDAC scheme. Second, the volume of the
fluid given by
\begin{equation}
  V_i = \frac{1}{\sum_j W_{ij}},
\end{equation}
is used in the discretization of the term $\frac{\nabla p}{\Varrho}$ whereas, in
PE-IPST-C the density $\Varrho$ is independent of neighbor particle positions. We
evaluate $\Varrho$ using a linear equation of state, \cref{eq:eos_invert}

In the EDAC scheme the initial configuration of particles affects the
results. Therefore, we consider an unperturbed configuration as shown in
\cref{fig:diff_pack}. In order to reduce the complexity, we consider an inviscid
MS given by
\begin{equation}
  \begin{split}
    u(x, y) =& \sin{\left(2 \pi x \right)} \cos{\left(2 \pi y \right)}\\
    v(x, y) =& - \sin{\left(2 \pi y \right)} \cos{\left(2 \pi x \right)}\\
    p(x, y) =& \cos{\left(4 \pi x \right)} + \cos{\left(4 \pi y \right)}.\\
  \end{split}
  \label{eq:mms_invis}
\end{equation}
Thus, the solver must maintain the pressure and velocity fields in the absence
of the viscosity. The source term for the EDAC scheme is given by
\begin{widetext}
\begin{equation}
  \begin{split}
    s_u(x, y) =& 2 \pi u \cos{\left(2 \pi x \right)} \cos{\left(2 \pi y \right)}
    - 2 \pi v \sin{\left(2 \pi x \right)} \sin{\left(2 \pi y \right)} - \frac{4
    \pi \sin{\left(4 \pi x \right)}}{\rho}\\
    s_v(x, y) =& 2 \pi u \sin{\left(2 \pi x \right)} \sin{\left(2 \pi y \right)}
    - 2 \pi v \cos{\left(2 \pi x \right)} \cos{\left(2 \pi y \right)} - \frac{4
    \pi \sin{\left(4 \pi y \right)}}{\rho}\\
    s_p(x, y) =& - 1.25 h \left(- 16 \pi^{2} \cos{\left(4 \pi x \right)} - 16 \pi^{2} \cos{\left(4
    \pi y \right)}\right) - 4 \pi u \sin{\left(4 \pi x \right)} - 4 \pi v \sin{\left(4
    \pi y \right)}.\\
  \end{split}
  \label{eq:source_invis_edac}
\end{equation}
\end{widetext}
We note that the source term employs density $\rho$ which is a function of
particle position given by $\frac{m_i}{V_i}$, where $m_i$ is the mass of the
particle.  In the case of the PE-IPST-C scheme, the source term is given by
\begin{widetext}
\begin{equation}
  \begin{split}
    s_u(x, y) =& 2 \pi u \cos{\left(2 \pi x \right)} \cos{\left(2 \pi y \right)}
    - 2 \pi v \sin{\left(2 \pi x \right)} \sin{\left(2 \pi y \right)} - \frac{4
    \pi \sin{\left(4 \pi x \right)}}{\Varrho}\\
    s_v(x, y) =&2 \pi u \sin{\left(2 \pi x \right)} \sin{\left(2 \pi y \right)}
    - 2 \pi v \cos{\left(2 \pi x \right)} \cos{\left(2 \pi y \right)} - \frac{4
    \pi \sin{\left(4 \pi y \right)}}{\Varrho}\\
    s_p(x, y)=& - 1.25 h \left(- 16 \pi^{2} \cos{\left(4 \pi x \right)} - 16 \pi^{2} \cos{\left(4
    \pi y \right)}\right) - 4 \pi u \sin{\left(4 \pi x \right)} - 4 \pi v \sin{\left(4
    \pi y \right)}.\\
  \end{split}
  \label{eq:source_invis_pe_ipst}
\end{equation}
\end{widetext}
We note that the source term $s_p$ in \cref{eq:source_invis_edac} and
\cref{eq:source_invis_pe_ipst} are same. We simulate the problem with the MS in
\cref{eq:mms_invis}. The (orange) solid boundary properties are reset using this
MS before every time step.

\begin{figure}[ht!]
  \centering
  \includegraphics[width=\linewidth]{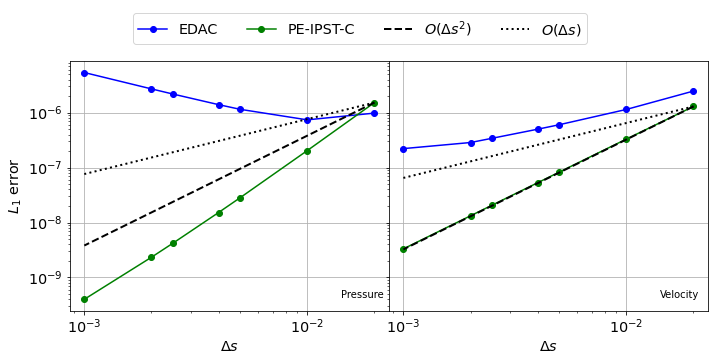}
  \caption{The error in pressure (left) and velocity (right) with fluid particles
  initialized using the MS in \cref{eq:mms_invis}, and the source term in
  \cref{eq:source_invis_edac} for EDAC and \cref{eq:source_invis_pe_ipst} for
  PE-IPST-C after 1 timestep.}
  \label{fig:diff_edac}
\end{figure}

\begin{figure}[ht!]
  \centering
  \includegraphics[width=\linewidth]{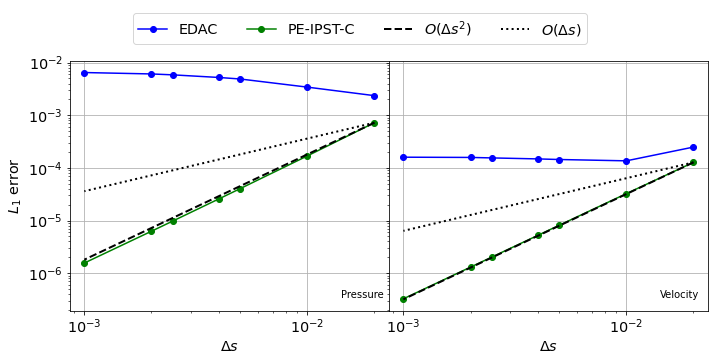}
  \caption{The error in pressure (left) and velocity (right) with fluid particles
  initialized using the MS in \cref{eq:mms_invis}, and the source term in
  \cref{eq:source_invis_edac} for EDAC and \cref{eq:source_invis_pe_ipst} for
  PE-IPST-C after 100 timestep.}
  \label{fig:diff_edac_100}
\end{figure}

In \cref{fig:diff_edac}, we plot the $L_1$ error in pressure and velocity after
one timestep for both the schemes. Clearly, the EDAC case diverges in the case
of pressure, whereas we observe a reduced order of convergence in velocity. In
contrast, the PE-IPST-C scheme shows second-order convergence in velocity and
higher in case of pressure. We observe this increased order only for the first
iteration. In \cref{fig:diff_edac_100}, we plot the $L_1$ error in pressure and
velocity after 100 timesteps for both the schemes. In the case of the EDAC scheme,
the order of convergence in the velocity does not remains first-order whereas,
the L-IPST-C scheme shows second-order convergence in both pressure and
velocity.

We note that, we use an unperturbed mesh therefore we must obtain second-order
convergence to the level of discretization error for 1 timestep in the case of
the EDAC scheme as well.  We observe this behavior since $\rho$ (a function of
neighbor particle positions) is present in the source term which comes from the
governing differential equation. Therefore, as mentioned in
\onlinecite{negi2021numerical}, we should treat $\rho$ as a separate property as we do
in the case of the PE-IPST-C scheme.

\subsection{Comparison of E-C and TV-C}
\label{sec:ec_tvc}

In this test case, we apply MMS to E-C and TV-C
schemes introduced in \cref{sec:sph}. The governing equations for E-C scheme is given
in \cref{eq:ec_eq2} whereas for TV-C in \cref{eq:wc_ale}. The expression for the
source terms turns out to be same for both \cref{eq:ec_eq2} and \cref{eq:wc_ale}
governing equations given by
\begin{equation}
  \begin{split}
    s_\Varrho = & \frac{\partial \Varrho}{\partial t} + \Varrho \nabla \cdot \ten{u} +
    \ten{u} \cdot \nabla \Varrho,\\
    s_\ten{u} = & \frac{\partial \ten{u}}{\partial t} + \frac{\nabla p}{\Varrho} - \nu \nabla^2 \ten{u} +
    \ten{u} \cdot \nabla \ten{u}.\\
  \end{split}
  \label{eq:s_tvc_ec}
\end{equation}
These source terms are the same as obtained in the case of the L-IPST-C scheme
as well. In E-C scheme, we fix the grid and add the convective term as the
correction, whereas in TV-C scheme, we add the shifting velocity in the LHS of
the governing equations.

In order to show the convergence of the scheme, we consider the inviscid MS in
\cref{eq:mms_invis} with the linear EOS. We do not consider the viscous term
since the term introduces similar error in both the schemes. We write the source
term as
\begin{widetext}
\begin{equation}
  \begin{split}
    s_u(x, y) =& 2 \pi u \cos{\left(2 \pi x \right)} \cos{\left(2 \pi y \right)}
    - 2 \pi v \sin{\left(2 \pi x \right)} \sin{\left(2 \pi y \right)} - \frac{4
    \pi \sin{\left(4 \pi x \right)}}{\Varrho},\\
    s_v(x, y) =& 2 \pi u \sin{\left(2 \pi x \right)} \sin{\left(2 \pi y \right)}
    - 2 \pi v \cos{\left(2 \pi x \right)} \cos{\left(2 \pi y \right)} - \frac{4
    \pi \sin{\left(4 \pi y \right)}}{\Varrho},\\
    s_\Varrho(x,y) = & - \frac{4 \pi u \sin{\left(4 \pi x \right)}}{c_{0}^{2}} -
    \frac{4 \pi v \sin{\left(4 \pi y \right)}}{c_{0}^{2}},\\
  \end{split}
  \label{eq:source_tvc_ec}
\end{equation}
\end{widetext}
where $\ten{s}_\ten{u}=s_u \hat{\ten{i}} + s_u \hat{\ten{j}}$ is the source term
for the momentum equation in both the schemes. We consider an unperturbed initial
particle distribution and run the simulation for 100 timesteps. The particles
are initialized with the MS in \cref{eq:mms_invis} and solid boundary are reset
using the MS before every time step.

\begin{figure}[ht!]
  \centering
  \includegraphics[width=\linewidth]{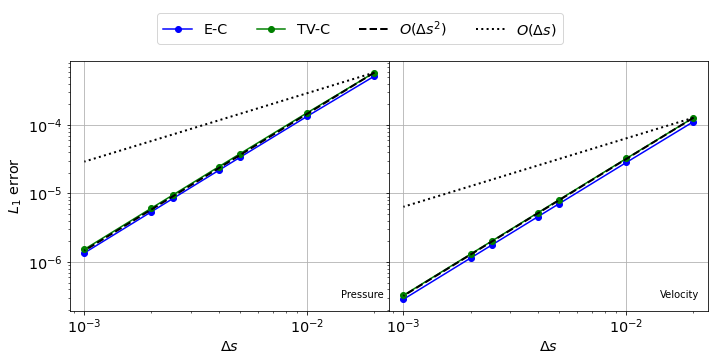}
  \caption{The error in pressure (left) and velocity (right) with fluid particles
  initialized using the MS in \cref{eq:mms_invis} and the source term in
  \cref{eq:source_tvc_ec} after 100 timesteps for the
  different schemes.}
  \label{fig:tvc_ec}
\end{figure}

In \cref{fig:tvc_ec}, we plot the $L_1$ error in pressure and velocity as a
function of resolution for both the schemes. Since we use second-order accurate
discretization in both the schemes, they show second-order convergence in both
pressure and velocity as expected. Thus, we see that the modified governing
equations (\cref{eq:wc_ale} and \cref{eq:ec_eq2}) must be considered to obtain
the source term for the schemes.

\subsection{Identification of mistakes in the implementation}
\label{sec:verify}

In this section, we demonstrate the use of MS as a technique to identify
mistakes in the implementation. We use the L-IPST-C scheme, and introduce
either erroneous or lower order discretization for a single term in the
governing equations. We then use the proposed MMS to identify the problem.

\subsubsection{Wrong divergence estimation}
We introduce an error in the discretized form of the continuity equation used in
the L-IPST-C scheme. We refer to this modified scheme as \emph{incorrect CE}. We
write the \emph{incorrect} discretization for the divergence of velocity as
\begin{equation}
  \left< \nabla \cdot \ten{u} \right> = \sum_j (\ten{u}_j \textcolor{red}{+}
  \ten{u}_i) \cdot \tilde{\nabla} W_{ij} \omega_j,
  \label{eq:wrong_ce}
\end{equation}
where the error is shown in red. Since only the continuity equation is involved,
we use the inviscid MS given by
\begin{equation}
  \begin{split}
    u(x, y) = & \left(y-1\right)^{2} \sin{\left(2 \pi x \right)} \cos{\left(2 \pi y \right)}\\
    v(x, y) = & - \sin{\left(2 \pi y \right)} \cos{\left(2 \pi x \right)}\\
    p(x, y) = & \left(y - 1\right) \left(\cos{\left(4 \pi x \right)} +
    \cos{\left(4 \pi y \right)}\right)\\
  \end{split}
  \label{eq:mms_invis_no_div}
\end{equation}
The source terms can be determined by subjecting the above MS to
\cref{eq:ns_wc}. We simulate the problem for 1 timestep with a packed domain
(see \cref{fig:diff_pack}). In order to test erroneous or lower order
discretization in the scheme, we recommend the simulation of only one timestep
with a packed initial particle distribution.
\begin{figure}[ht!]
  \centering
  \includegraphics[width=\linewidth]{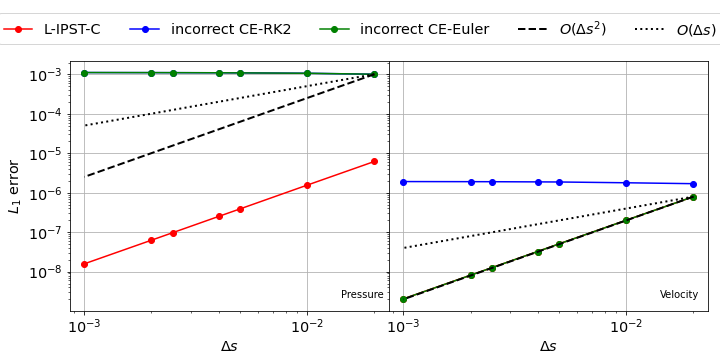}
\caption{The error in pressure (left) and velocity (right) with fluid particles
initialized using the MS in \cref{eq:mms_invis} and the source term in
\cref{eq:source_tvc_ec} after 1 timestep for L-IPST-C and the scheme with the
divergence computed using the incorrect \cref{eq:wrong_ce}.}
  \label{fig:wrong_ce}
\end{figure}

In \cref{fig:wrong_ce}, we plot the $L_1$ error in pressure and velocity as
a function of the resolution for the L-IPST-C scheme and its variant
\emph{incorrect CE} with two time integrators, Euler and RK2.  Clearly, the
error in pressure increases by a significant amount and the order of
convergence is zero for \emph{incorrect CE}. However, the error in pressure
propagates to velocity in case of the RK2 integrator. Therefore, we
recommend that one use single stage integrators while using MMS as a
technique to identify mistakes. By looking at \emph{incorrect CE-Euler}
plot in \cref{fig:wrong_ce} we can immediately infer that there is an error
in either the continuity equation or the equation of state.

\subsubsection{Using a symmetric pressure gradient discretization}
In this testcase, we use a symmetric
formulation as used by \onlinecite{monaghan-review:2005,sun2017deltaplus,negi2021numerical} for
the pressure gradient term in the L-IPST-C scheme. We refer to this method as
\emph{sym}. Since only the pressure gradient is involved, we use the same MS as
in the previous case.
\begin{figure}[ht!]
  \centering
  \includegraphics[width=\linewidth]{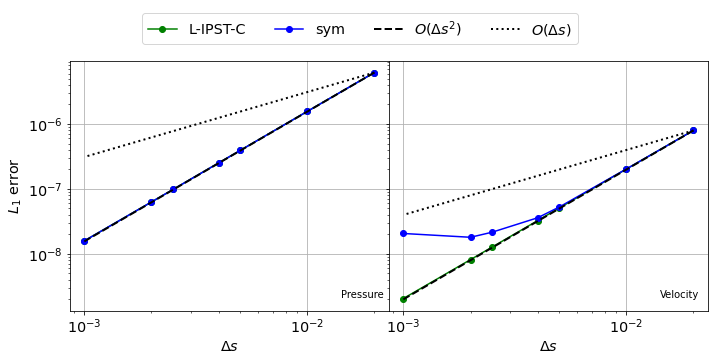}
  \caption{The error in pressure (left) and velocity (right) with fluid particles
  initialized using the MS in \cref{eq:mms_invis} and the source term in
  \cref{eq:source_tvc_ec} after 1 timestep for L-IPST-C and the scheme
  with pressure gradient computed using symmetric formulation.}
  \label{fig:wrong_pe}
\end{figure}

In \cref{fig:wrong_pe}, we plot the $L_1$ error after 1 timestep in pressure and
velocity as a function of resolution for L-IPST-C and \emph{sym} schemes.
Clearly, the order of convergence is affected in the velocity only. Therefore,
it is evident that a inconsistent pressure gradient discretization is used.

\subsubsection{Using inconsistent discrete viscous operator}
In this testcase, we use the formulation proposed by
\citet{cleary1999conduction} to approximate the viscous term in the L-IPST-C
scheme. We refer to this method as \emph{Cleary}. Since viscosity is involved,
we use the MS involving viscous effect given by \cref{eq:mms_vis_no_div}. While
testing the viscous term we use a high value of $\nu=.25m^2/s$ such that the
error due to viscosity dominates the error in the momentum equation. We simulate
the problem with a packed configuration of particles for 1 timestep using the MS
in \cref{eq:mms_vis_no_div} and with the corresponding source terms. We fix the
timestep using \cref{eq:dt} such that we satisfy the stability condition.
\begin{figure}[ht!]
  \centering
  \includegraphics[width=\linewidth]{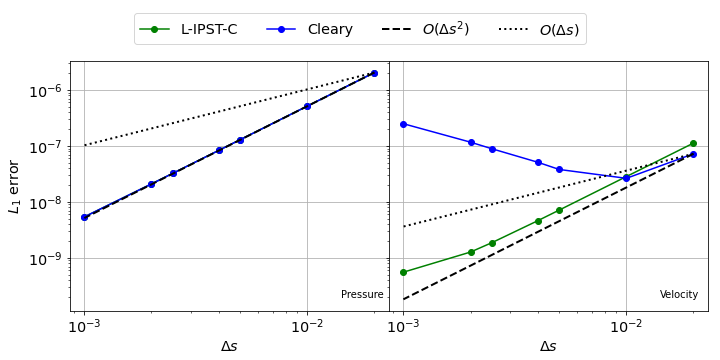}
\caption{The error in pressure (left) and velocity (right) with fluid particles
initialized using the MS in \cref{eq:mms_vis_no_div} and the corresponding source term
after 1 timestep for L-IPST-C and the scheme with viscous term
discretized using formulation given by \citet{cleary1999conduction}.}
  \label{fig:wrong_ve}
\end{figure}

In \cref{fig:wrong_ve}, we plot the $L_1$ error in pressure and velocity as a
function of resolution for L-IPST-C and \emph{Cleary} schemes. Since the viscous
formulation by \citet{cleary1999conduction} does not converge in the perturbed
domain~\cite{negi2021numerical}, we observe divergence in the velocity.
Therefore, we infer that there is an error in the viscous term.

\subsection{MMS applied to boundary condition}
\label{sec:bc}

\begin{figure}[ht!]
  \centering
  \includegraphics[width=0.7\linewidth]{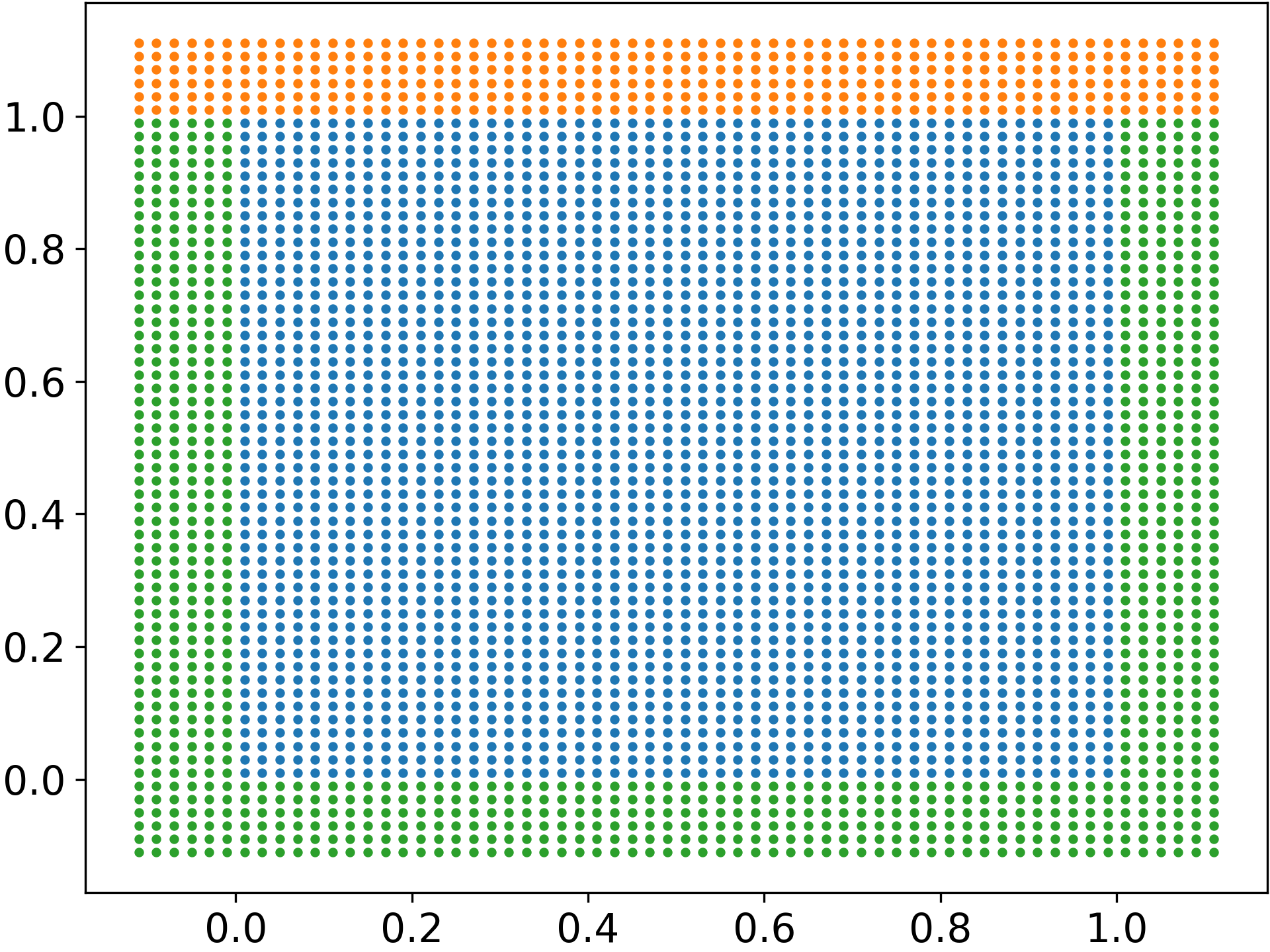}
  \caption{Different particle used for testing the boundary condition with fluid in blue, MS solid boundary in
  green, and SPH solid boundary in orange.}
  \label{fig:solid_pack}
\end{figure}

In this section, we use MMS to verify the convergence of boundary conditions in
SPH. In order to do this, the scheme used must converge at least as fast as the
boundary conditions. Therefore, we consider the second-order convergent L-IPST-C
scheme. We study the Dirichlet boundary conditions for pressure and velocity,
no-slip and slip velocity boundary conditions, and the Neumann pressure
boundary condition. We consider an unperturbed domain as shown in
\cref{fig:solid_pack}, where we solve the fluid equations using the L-IPST-C
scheme for the blue particles and set the MS before every time step for the
green particles. We set the properties in the orange particles using the
appropriate boundary condition we intend to test. For example, if we set the
pressure Dirichlet boundary condition in SPH then we set velocity and density
using the MS.
In order to obtain rate of convergence, we evaluate $L_\infty$ error using,
\begin{equation}
  L_\infty(N) = max\{|f(\ten{x}_i) - f(\ten{x}_o)|, i = 1,\dots, N\},
  \label{eq:linf}
\end{equation}
where N is the total number of fluid particles for which $y>0.9$, and
$f(\ten{x}_i)$ and $f(\ten{x}_o)$ are the computed and exact value of the
property of interest, respectively. We consider only a portion near the
boundary since only that region is affected the most by the boundary
implementation. In the following sections, we test the different boundary
conditions in SPH using MMS.

\subsubsection{Dirichlet boundary condition}

In this testcase, we construct the MS for boundary condition as discussed in
\cref{sec:mms}. In order to set the homogenous boundary condition at $y=1$, we
modify the MS in \cref{eq:mms_invis} as
\begin{equation}
  \begin{split}
    u =& \left(y - 1\right) \sin{\left(2 \pi x \right)} \cos{\left(2 \pi y
    \right)}\\
    v = &- \left(y - 1\right) \sin{\left(2 \pi y \right)} \cos{\left(2 \pi x
    \right)}\\
    p=&\left(y - 1\right) \left(\cos{\left(4 \pi x \right)} + \cos{\left(4 \pi y
    \right)}\right)\\
  \end{split}
  \label{eq:mms_dirichlet}
\end{equation}
Clearly, at $y=1$ we have boundary values $u=v=p=0$. In SPH, the Dirichlet
boundary may be applied by setting the desired value of the property on the
ghost layer shown in orange in \cref{fig:solid_pack}. We set homogenous velocity
and pressure boundary conditions in two separate testcases and refer to them as
\emph{velocity BC} and \emph{pressure BC}, respectively. We set the
pressure/velocity on the solid using the MS when we set velocity/pressure using
the SPH method. We simulate the problem for 100 timesteps with the MS in
\cref{eq:mms_dirichlet}.
\begin{figure}[ht!]
  \centering
  \includegraphics[width=\linewidth]{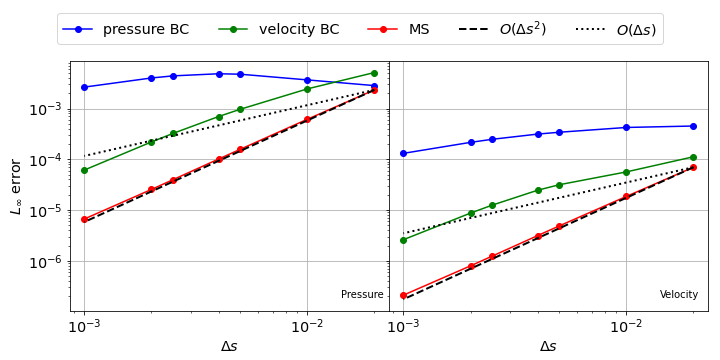}
  \caption{The error in pressure (left) and velocity (right) with fluid particles
  initialized using the MS in \cref{eq:mms_dirichlet} 100 timesteps for
  L-IPST-C and \emph{velocity BC} and \emph{pressure BC} applied at the orange boundary
  in \cref{fig:solid_pack}.}
  \label{fig:dirc}
\end{figure}

In \cref{fig:dirc}, we plot the $L_\infty$ error in pressure and velocity as a
function of resolution for L-IPST-C, \emph{velocity BC}, and \emph{pressure BC}.
Clearly, both the boundary conditions introduce error in the solution. The error
introduced due to \emph{Velocity BC} remains around second-order in pressure and
first-order in velocity. The \emph{pressure BC} is rarely used in SPH and
introduces a significant amount of error with almost zero order convergence.

\subsubsection{Slip boundary condition}

In the SPH method, the slip boundary condition can be applied using the method
proposed by \citet{maciaTheoreticalAnalysisNoSlip2011}. First, we extrapolate the
velocity of the fluid to the solid using
\begin{equation}
  \ten{u}_s = \frac{\sum \ten{u}_f W_{sf}}{\sum_j W_{sf}},
  \label{eq:shep_vel}
\end{equation}
where $\ten{u}_s$ and $\ten{u}_f$ denotes the velocity of wall and fluid
particles, respectively. Then, we reverse the component of the velocity normal
to the wall. This method ensures that the divergence of velocity is captured
accurately near the slip wall. Therefore, we consider the inviscid MS given by
\begin{equation}
  \begin{split}
    u(x, y) =& \left(y - 1\right)^{2} \sin{\left(2 \pi x \right)} \cos{\left(2
    \pi y \right)}\\
    v(x, y) = &- \sin{\left(2 \pi y \right)} \cos{\left(2 \pi x
    \right)}\\
    p(x, y)=&\left(y - 1\right) \left(\cos{\left(4 \pi x \right)} + \cos{\left(4 \pi y
    \right)}\right)\\
  \end{split}
  \label{eq:mms_slip}
\end{equation}
We note that the $u$ velocity is symmetric across $y=1$ and $v$ velocity is
asymmetric. We consider the domain as shown in \cref{fig:solid_pack} and apply
the free slip boundary condition on the solid boundary shown in orange color for
the L-IPST-C scheme. We refer to this method as \emph{slip BC}. We note that the
pressure and density on the solid is set using the MS. We simulate the
problem for 100 timesteps.
\begin{figure}[ht!]
  \centering
  \includegraphics[width=\linewidth]{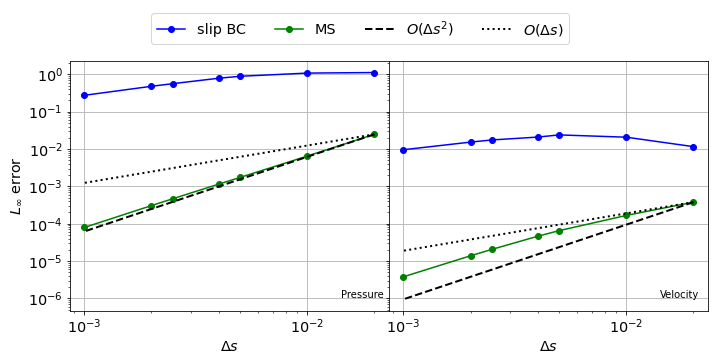}
  \caption{The error in pressure (left) and velocity (right) with fluid particles
  initialized using the MS in \cref{eq:mms_slip} after 100 timesteps for
  L-IPST-C and \emph{slip BC} applied on the orange boundary
  in \cref{fig:solid_pack}.}
  \label{fig:slip}
\end{figure}
In \cref{fig:slip}, we plot the $L_1$ error in pressure and velocity as a
function of resolution for L-IPST-C and \emph{slip BC} schemes. Clearly, the
application of slip boundary condition increases the error and the order of
convergence is less than one. In the case of the L-IPST-C scheme, the lower
resolutions show first order convergence but as the resolution increases
approaches second-order. We note that the \cref{fig:slip} shows the $L_\infty$
error, however convergence of the $L_1$ error is close to second-order for all
resolutions. In summary, the slip boundary condition as proposed in
\onlinecite{maciaTheoreticalAnalysisNoSlip2011} is accurate in velocity but reduces
the accuracy of the pressure.

\subsubsection{Pressure boundary condition}

In the pressure boundary condition proposed by
\citet{maciaTheoreticalAnalysisNoSlip2011}, we ensure that the pressure gradient
normal to the boundary is zero. We apply the boundary condition by setting the
pressure of the solid boundary particles using
\begin{equation}
  p_s = \frac{\sum p_f W_{sf}}{\sum_j W_{sf}},
\end{equation}
where $p_s$ and $p_f$ denotes the pressure of wall and fluid particles,
respectively. For simplicity, we ignore the acceleration due to gravity and
motion of the solid body. We consider the MS of the form
\begin{equation}
  \begin{split}
    u(x, y) =& y^{2} \sin{\left(2 \pi x \right)} \cos{\left(2 \pi y \right)}\\
    v(x, y) = &- \sin{\left(2 \pi y \right)} \cos{\left(2 \pi x
    \right)}\\
    p(x, y)=&\left(y - 1\right)^2 \left(\cos{\left(4 \pi x \right)} + \cos{\left(4 \pi y
    \right)}\right)\\
  \end{split}
  \label{eq:mms_pressure}
\end{equation}
Clearly, the MS satisfies $\frac{\partial p}{\partial y}=0$ at $y=1$. We consider the domain as
shown in \cref{fig:solid_pack} and apply the pressure boundary condition on the
solid boundary shown in orange color for L-IPST-C scheme. We refer to this
method as \emph{Neumann BC}. We simulate the problem for 100 timesteps.
\begin{figure}[ht!]
  \centering
  \includegraphics[width=\linewidth]{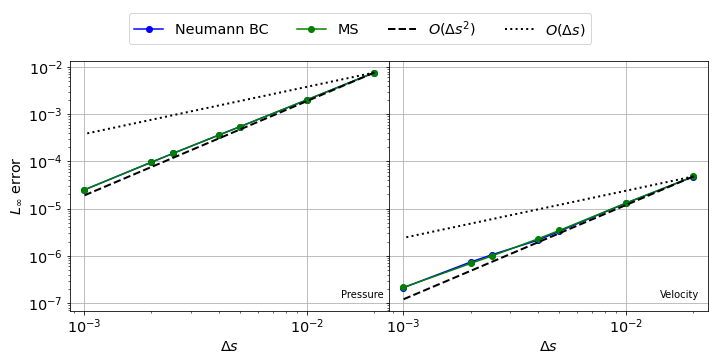}
  \caption{The error in pressure (left) and velocity (right) with fluid particles
  initialized using the MS in \cref{eq:mms_pressure} after 100 timesteps for
  L-IPST-C and \emph{Neumann BC} applied on the orange boundary
  in \cref{fig:solid_pack}.}
  \label{fig:pres_bc}
\end{figure}

In \cref{fig:pres_bc}, we plot the $L_\infty$ error in pressure and velocity for
L-IPST-C and \emph{Neumann BC}. The results show that the pressure boundary
condition is second order convergent.

\subsubsection{No-slip boundary condition}
\citet{maciaTheoreticalAnalysisNoSlip2011} proposed the no-slip boundary
condition for SPH where we set the wall velocity as
\begin{equation}
  \ten{u}_s = 2  \ten{u}_w - \ten{\tilde{u}}_s,
\end{equation}
where $\ten{u}_w$ is velocity of the wall and $\ten{\tilde{u}}_s$ is the Shepard
interpolated velocity (see \cref{eq:shep_vel}). In the no-slip boundary, we
ensure that $\frac{\partial u}{\partial y} = 0$ at $y=1$ therefore, we use the
MS for viscous flow given by
\begin{equation}
  \begin{split}
    u(x, y, t) = & \left(y - 1\right)^{2} e^{- 10 t} \sin{\left(2 \pi x \right)}
    \cos{\left(2 \pi y \right)}\\
    v(x, y, t) = & - \left(y - 1\right)^{2} e^{- 10 t} \sin{\left(2 \pi y \right)}
    \cos{\left(2 \pi x \right)}\\
    p(x, y, t) = & \left(\cos{\left(4 \pi x \right)} + \cos{\left(4 \pi y
    \right)}\right) e^{- 10 t}\\
  \end{split}
  \label{eq:mms_no_slip}
\end{equation}

We consider the domain as shown in \cref{fig:solid_pack} and apply the pressure
boundary condition on the solid boundary shown in orange color for the L-IPST-C
scheme. We refer to this method as \emph{no-slip BC}. We simulate the problem
for 100 timesteps with $\nu=1.0m^2/s$.
\begin{figure}[ht!]
  \centering
  \includegraphics[width=\linewidth]{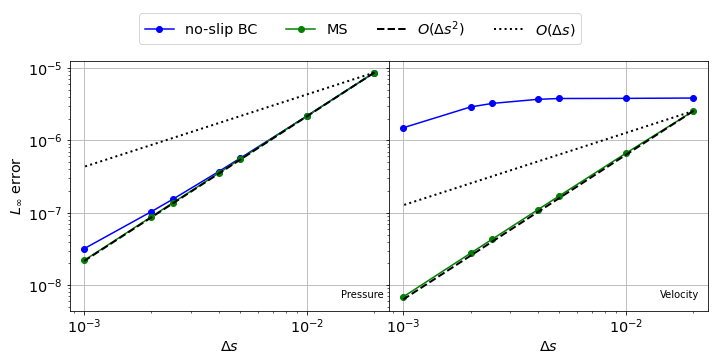}
  \caption{The error in pressure (left) and velocity (right) with fluid particles
  initialized using the MS in \cref{eq:mms_pressure} after 100 timesteps for
  L-IPST-C and \emph{no-slip BC} applied on the orange boundary in
  \cref{fig:solid_pack}.}
  \label{fig:no_slip_bc}
\end{figure}

In \cref{fig:no_slip_bc}, we plot the $L_\infty$ error in pressure and velocity
for 100 timesteps. Clearly, the \emph{no-slip BC} shows increased error and a
zero-order convergence. However, it does not introduce error in the pressure.

Thus in this section, we have demonstrated the MMS for obtaining the order of
convergence of boundary condition implementations in SPH.

\subsection{Convergence and extreme resolutions}
\label{sec:extreme}

\begin{figure}[htbp]
  \centering
  \includegraphics[width=0.7\linewidth]{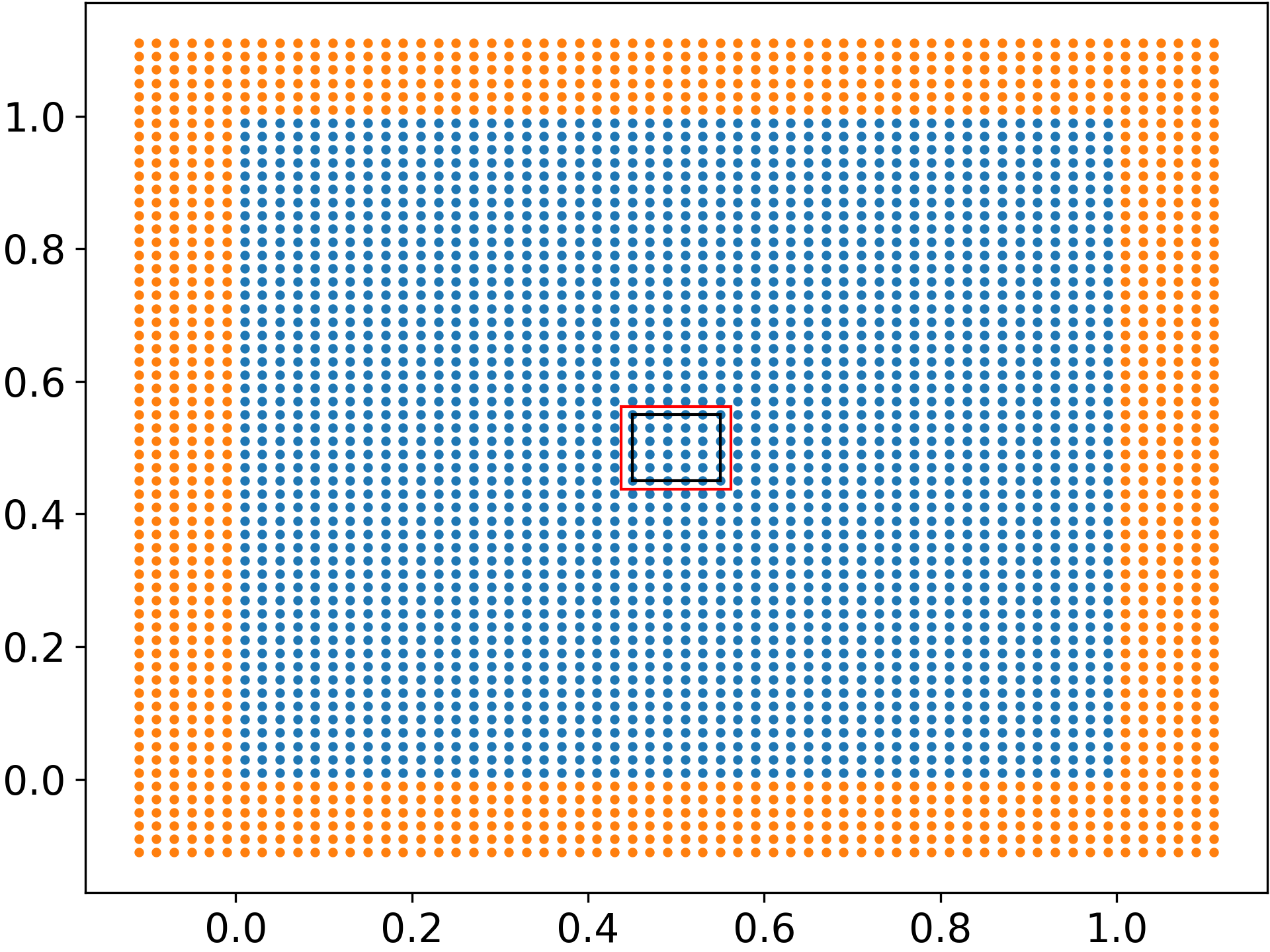}
  \caption{The domain filled by blue fluid particles. The red box shows the
  smallest domain considered for the highest resolution of $8000 \times 8000$
  and the black box shows the area which is considered to evaluate error for all
  the resolutions.}
  \label{fig:var_domain}
\end{figure}

Thus far we have used particle resolutions in the range $10^{-3} \le \Delta s
\le 2\times 10^{-2}$. We wish to study the convergence of the scheme when much
higher resolutions are considered. We consider a domain of size $1\times1$ with
uniformly distributed particles as shown in \cref{fig:var_domain}. In order to
reduce computation, we reduce the size of the domain by half if the number of
particles crosses $1M$. In the \cref{fig:var_domain}, the red box shows the
domain considered for the computation which one million particles with $\Delta s
= 1.25 \times 10^{-4}$. In order to obtain an unbiased error estimate we
consider same MS and the domain shown by black box in \cref{fig:var_domain} to
evaluate $L_\infty$ error using \cref{eq:linf}.

We first consider the MS given in \cref{eq:mms_vis_no_div}. We solve the
\cref{eq:ns_wc_mod} using the L-IPST-C scheme for all the resolutions with
$\nu=.01m^2/s$. We consider the case where we do not correct the kernel gradient
in the discretization of \cref{eq:ns_wc_mod} in the L-IPST-C scheme.

\begin{figure}[htbp]
  \centering
  \includegraphics[width=\linewidth]{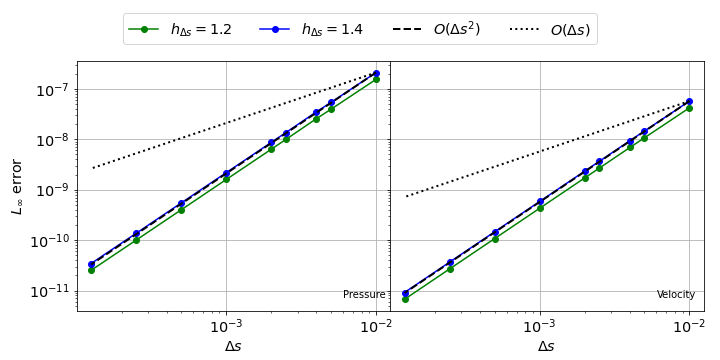}
\caption{The error in pressure (left) and velocity (right) as a function of
resolution for two different $h_{\Delta s}$ values with the MS in
\cref{eq:mms_vis_no_div}. All cases are solved using L-IPST-C scheme with kernel
gradient correction.}
  \label{fig:lim_mms1}
\end{figure}

\begin{figure}[htbp]
  \centering
  \includegraphics[width=\linewidth]{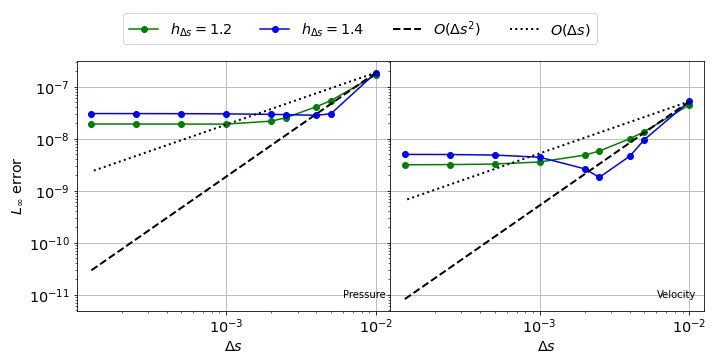}
\caption{The error in pressure (left) and velocity (right) as a function of
resolution for two different $h_{\Delta s}$ values with the MS in
\cref{eq:mms_vis_no_div}. All cases are solved using L-IPST-C scheme with no kernel
gradient correction.}
  \label{fig:lim_mms1_nc}
\end{figure}

In \cref{fig:lim_mms1}, we plot the error in pressure and velocity solved using
L-IPST-C scheme with kernel gradient corrected, after 100 timesteps as a
function of resolution for $h_{\Delta s}=1.2$ and $h_{\Delta s}=1.4$. Clearly,
We obtain second order convergence. In
\cref{fig:lim_mms1_nc}, we plot the error for the case where we do not employ
kernel gradient correction. Clearly, the discretization error dominates.

We also consider the MS containing a range of frequencies given by
\begin{equation}
  \begin{split}
    u(x, y, t) =& y^2 e^{- 10 t} \sum_{j=1}^{10} \sin{\left(2 j \pi x \right)} \cos{\left(2 j \pi y \right)}\\
    v(x, y, t) =& - e^{- 10 t} \sum_{j=1}^{10} \sin{\left(2 j \pi y \right)} \cos{\left(2 j \pi x \right)}\\
    p(x, y, t) =&  e^{- 10 t} \sum_{j=1}^{10} \cos{\left(4 j \pi x \right)} + \cos{\left(4 j \pi y \right)}.\\
  \end{split}
  \label{eq:lim_mms9}
\end{equation}
We simulate the \cref{eq:ns_wc} using L-IPST-C scheme for the above MS. As
before, we also consider the case where we do not employ kernel correction.

\begin{figure}[htbp]
  \centering
  \includegraphics[width=\linewidth]{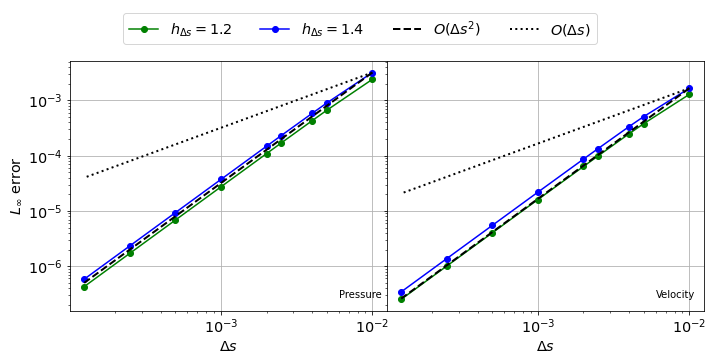}
  \caption{The error in pressure (left) and velocity (right) as a function of
  resolution for two different $h_{\Delta s}$ values with the MS in
  \cref{eq:lim_mms9}. All cases are solved using L-IPST-C scheme with kernel
  gradient correction.}
  \label{fig:lim_mms9}
\end{figure}

\begin{figure}[htbp]
  \centering
  \includegraphics[width=\linewidth]{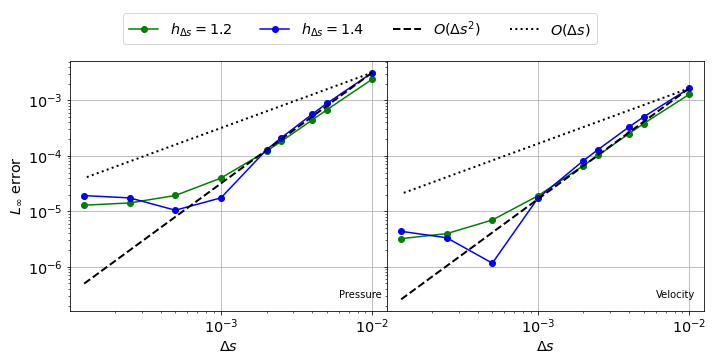}
  \caption{The error in pressure (left) and velocity (right) as a function of
  resolution for two different $h_{\Delta s}$ values with the MS in
  \cref{eq:lim_mms9}. All cases are solved using L-IPST-C scheme with no kernel
  gradient correction.}
  \label{fig:lim_mms9_nc}
\end{figure}

In \cref{fig:lim_mms9}, we plot the error in pressure and velocity solved using
L-IPST-C scheme with kernel gradient correction for 100 timesteps as a function
of resolutions. Clearly, both the cases shows second-order convergence. In
\cref{fig:lim_mms9_nc}, we plot the error in pressure and velocity for the
solution obtained using L-IPST-C scheme with no kernel correction. As can be
seen the kernel correction is essential in order to obtain second-order
convergence at high resolutions.

We have therefore shown that we can consider very high resolutions using the MMS
technique. This enables us to find flaws in the scheme which may not converge at
very high resolution. These are hard to test using traditional methods where an
actual problem is solved.

\subsection{Verification in 3D}
\label{sec:3d}

We now use the MMS to verify a three dimensional solver. Since the number of
particles in three-dimensions increase much faster than in two-dimensions, we
can reduce the domain size with resolution as done while dealing with extreme
resolutions. We consider a unit cube domain size with 1 million particles. As
we increase the resolution, we decrease the size of the domain such that the
number of particles in the domain remains at 1 million. We consider the MS given
by
\begin{equation}
  \begin{split}
    u(x, y, z, t) = &y^{2} e^{- 10 t} \sin{\left(\pi \left(2 x + 2 z\right) \right)} \cos{\left(\pi
      \left(2 x + 2 y\right) \right)}\\
    v(x, y, z, t) = &- e^{- 10 t} \sin{\left(\pi \left(2 y + 2 z\right) \right)} \cos{\left(\pi
      \left(2 x + 2 y\right) \right)}\\
    w(x, y, z, t) = &- e^{- 10 t} \sin{\left(\pi \left(2 x + 2 z\right) \right)} \cos{\left(\pi
      \left(2 y + 2 z\right) \right)}\\
    p(x, y, z, t) = & \left(\cos{\left(\pi \left(4 x + 4 y\right) \right)} + \cos{\left(\pi \left(4
    x + 4 z\right) \right)}\right) e^{- 10 t}.\\
  \end{split}
  \label{eq:3d_mms}
\end{equation}
We obtain the source term by subjecting the MS in \cref{eq:3d_mms} to the
governing equation in \cref{eq:ns_wc} with $\nu=0.01m^2/s$. We simulate the
problem for 10 timesteps.

\begin{figure}[htbp]
  \centering
  \includegraphics[width=\linewidth]{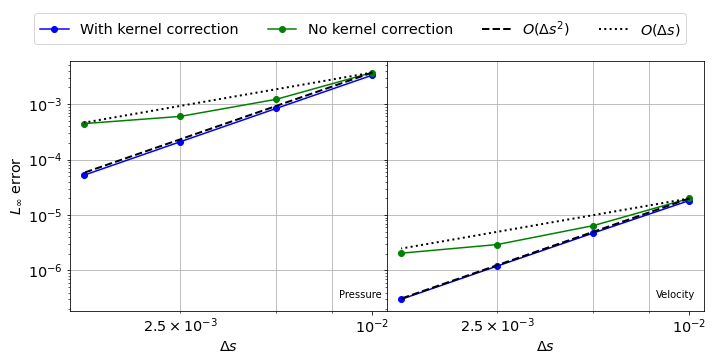}
\caption{The $L_\infty$ error in pressure (left) and velocity (right) after 10 timesteps as
a function of resolution solved using L-IPST-C scheme with and without kernel
correction. The source term are calculated using the MS in \cref{eq:3d_mms}.}
  \label{fig:3d_mms}
\end{figure}

In \cref{fig:3d_mms}, we plot the $L_\infty$ error in pressure and velocity as
a function of resolution for L-IPST-C scheme with and without kernel
correction. As expected, the case with no kernel correction gradually flatten
due dominance of discretization error. The case with kernel correction shows
second order convergence in both pressure and velocity. Thus we see that we
can easily test the SPH method in a three-dimensional domain using the MMS.

\section{Discussion}
\label{sec:discuss}

We have used the MMS to verify the convergence of different WCSPH schemes.
Thus far, most of the numerical studies of the accuracy and convergence of the
WCSPH method have used either an exact solution like the Taylor-Green vortex
problem, or with an established solver, or experimental result. These methods
are therefore limited in their ability to detect specific problems in an SPH
implementation. This is true even in the recent work of
\citet{negi2021numerical} where a Taylor-Green problem and a Gresho-Chan
vortex problem is used. These are complex problems and obtaining a solution to
these involves a significant amount of computation. Moreover, if the results
do not produce the expected accuracy or convergence, the researcher does not
obtain much insight into the origin of the problem. Furthermore, the
established approaches do not offer any means to study the accuracy of
boundary condition implementations.

In this context, the proposed approach offers a multitude of advantages listed
and discussed below:
\begin{itemize}
\item The method is highly efficient in terms of execution time. We are
able to detect problems in the implementations of specific discretization
operators in less than 100 iterations. Even for our most challenging cases
with a million particles, the typical run time for a single computation on
a multi-core CPU does not exceed a few minutes. On the other hand, the
comparison study for the lid-driven cavity case in \cref{sec:sph_conv} took
150 minutes for the $200 \times 200$ resolution.
\item The method easily works in three dimensions and we demonstrate its
  applicability for a simple three-dimensional case. This is significant
  because traditional SPH verifications only use two-dimensional problems.
\item We can effectively test the boundary condition implementations through
  this method. In this work we have demonstrated this for Dirichlet and
  Neumann boundary conditions in both pressure and velocity.
\item The method allows us to identify very specific problems with a solver.
  Through a judicious choice of MS and time integrator, we can identify if the
  implementation of a specific governing equation is the source of a problem.
  We have demonstrated this with several examples in the preceding sections.
\item We are able to verify the order of convergence efficiently even for very
  high resolutions and thereby test if the scheme is truly second order
  convergent as the resolution increases. In the present work we have
  demonstrated this for extremely high resolutions (involving $8000 \times
  8000$ particles) without needing to simulate the problem for a long duration
  and also limiting the number of computational particles to a smaller number.
\item The method will work on any manufactured solution and this allows us to
  test the scheme with functions involving a large range of frequencies. In
  contrast, many exact solutions involve simple functional forms. Therefore by
  using the MMS the solver can be tested with a more challenging class of
  problems.
\end{itemize}

As a result of these significant advantages, the proposed method offers a
robust, efficient, and powerful method to verify the accuracy and convergence
of SPH schemes.

\section{Conclusions}
\label{sec:conclusions}

In this paper we propose the use of the method of manufactured solutions
(MMS) in order to verify an SPH solver. While the MMS technique is well
established in the context of mesh-based
methods~\cite{roache1998verification}, to the best of our knowledge it does
not appear to have been employed in the context of Lagrangian SPH schemes
thus far. The application of MMS to Lagrangian SPH method is non-trivial as
the particles move.

In the present work we show for the first time how the method can be
employed to verify the accuracy of any modern weakly-compressible SPH
scheme. Specifically, we note that for successful application of the MMS,
quantities like gradient of velocity should be evaluated using the scheme
and not with the gradient of the MS. In this paper, we apply PST to
restrict the particles to remain inside the domain boundaries allowing us
to apply MMS to arbitrary shaped boundaries without the need for addition
and deletion of particles. We compare different initial particle
distributions used in SPH to obtain a minimum number of iterations required
for a result independent of initial distribution. We also show that one
should not use a divergence free velocity field while using MMS in SPH for
verification. We compare the EDAC and the PE-IPST-C schemes and show that
the density should be used as a property independent of the neighbor
particle distribution. We show that the method works in arbitrary number of
dimensions, allows us to systematically identify problems quickly in
specific discretizations employed by the scheme, and makes it possible to
verify the accuracy of boundary condition implementations as well. We also
demonstrate that the recently proposed family of second order convergent
WCSPH schemes~\cite{negi2021numerical} are indeed second order accurate.
Finally, our implementation is open source
(\url{https://gitlab.com/pypr/mms_sph}) and our numerical experiments and
results presented are fully automated in the interest of reproducibility.
Given that convergence and accuracy of SPH schemes is a grand-challenge
problem in the SPH community~\cite{vacondio_grand_2020}, the present work
offers a valuable contribution.

In the future, we propose to use this method to study the accuracy and
convergence of the method in the context of the various solid boundary
conditions proposed in SPH. Using the method in the context of inlet and
outlet boundary conditions and for free-surfaces may prove challenging and
remain to be explored. The method may also be applied in the context of
incompressible SPH, compressible SPH, and multi-phase SPH schemes. We plan to
explore these problems in the future.

\section*{Acknowledgments}

Thanks to Dr.~Kadambari Devarajan for helping wordsmith the title.

\bibliography{references}

\end{document}